\documentclass[prd,nofootinbib,preprint,superscriptaddress]{revtex4}
\pdfoutput=1
\usepackage[T1]{fontenc}
\usepackage{amsmath,amssymb}
\usepackage{epsfig}
\usepackage{subfigure}
\usepackage{float}
\usepackage{graphicx}
\usepackage{bigints}
\usepackage[usenames,dvipsnames]{color}
\usepackage{slashed}
\usepackage{multirow}
\usepackage[colorlinks,citecolor=blue]{hyperref}
\usepackage{pdfpages}
\usepackage{color}
\usepackage{comment}
\usepackage{soul}
\newcommand{\ogw}{\Omega_\text{gw}}

\newcommand{\mhsqr}[1]{M_{H_1}^2}

\DeclareUnicodeCharacter{2212}{-}
\begin{document}
\title{Thermalised dark radiation in the presence of PBH: ${\rm \Delta N_{\rm eff}}$ and gravitational waves complementarity}

%%%%%%%%%   Authors   %%%%%%%%%%%%

\author{Nayan Das}
\email{nayan.das@iitg.ac.in}
\affiliation{Department of Physics, Indian Institute of Technology Guwahati, Assam 781039, India}

\author{Suruj Jyoti Das}
\email{suruj@iitg.ac.in}
\affiliation{Department of Physics, Indian Institute of Technology Guwahati, Assam 781039, India}

\author{Debasish Borah}
\email{dborah@iitg.ac.in}
\affiliation{Department of Physics, Indian Institute of Technology Guwahati, Assam 781039, India}

\begin{abstract}
We study the possibility of detecting dark radiation (DR) produced by a combination of interactions with the thermal bath and ultra-light primordial black hole (PBH) evaporation in the early universe. We show that the detection prospects via cosmic microwave background (CMB) measurements of the effective relativistic degrees of freedom ${\rm \Delta N_{eff}}$ get enhanced in some part of the parameter space compared to the purely non-thermal case where DR is produced solely from PBH. On the other hand, for certain part of the parameter space, DR which initially decouples from the bath followed by its production from PBH evaporation, can re-enter the thermal bath leading to much tighter constraints on the PBH parameter space. We also discuss the complementary detection prospects via observation of stochastic gravitational wave (GW) sourced by PBH density perturbations. The complementary probes offered by CMB and GW observations keep the detection prospects of such light degrees of freedom very promising in spite of limited discovery prospects at particle physics experiments.
\end{abstract}
%\pacs{}
\maketitle
%%%%%%%%%%%%%%%%%%%%%%%%%%%%%%%%%%%%%%%%%%%%%%%%%%%%%%%%%%%%%%%%%
\section{Introduction}
\label{sec:Intro}
The matter content in the present universe is dominated by dark matter (DM) as suggested by numerous astrophysics and cosmology based observations \cite{Zyla:2020zbs, Aghanim:2018eyx}. While a fundamental particle with the required characteristics can give rise to DM, the dark sector in general can be much richer. For example, the dark sector can contain different light degrees of freedom, commonly referred to as dark radiation (DR) \cite{Ackerman:2008kmp}. While their contribution to the overall energy budget of the present universe is negligible, they can have different phenomenological significance as well as detection prospects. For example, a very light scalar or gauge boson can give rise to sizeable self-interaction of DM having the potential to solve the small-scale structure issues of cold dark matter paradigm \cite{Spergel:1999mh, Tulin:2017ara, Bullock:2017xww}. Other relativistic degrees of freedom may arise in different particle physics scenarios in the form of (pseudo) Goldstone boson, moduli fields, graviton, light sterile neutrinos, Dirac active neutrinos and so on. 

Similar to radiation energy density at present epoch, these light relativistic degrees of freedom typically contributes negligibly to the total energy budget. In spite of that, cosmological observations can tightly constrain their abundance. Presence of dark radiation can be probed at cosmic microwave background (CMB) experiments. Existing data from CMB experiments like Planck constrain such additional light species by putting limits on the effective
degrees of freedom for neutrinos during the
era of recombination ($z\sim 1100$) as  \cite{Aghanim:2018eyx} 
\begin{eqnarray}
{\rm
N_{\rm eff}= 2.99^{+0.34}_{-0.33}
}
\label{Neff}
\end{eqnarray}
at $2\sigma$ or $95\%$ CL including baryon acoustic oscillation (BAO) data. At $1\sigma$ CL it becomes more stringent to ${\rm N}_{\rm eff} = 2.99 \pm 0.17$. Similar bound also exists from big bang nucleosynthesis (BBN) $2.3 < {\rm N}_{\rm eff} <3.4$ at $95\%$ CL \cite{Cyburt:2015mya}. All these bounds are consistent with SM predictions ${\rm N^{SM}_{eff}}=3.044$ \cite{Bennett:2020zkv, Froustey:2020mcq, Akita:2020szl}. (Note that references \cite{Mangano:2005cc, Grohs:2015tfy,deSalas:2016ztq, EscuderoAbenza:2020cmq} report slightly higher value of $N_{\rm eff}^{\rm SM}$ as 3.045.)\footnote{A very recent paper \cite{Cielo:2023bqp} reports $N_{\rm eff}^{\rm SM}=3.043$ by taking into account of the NLO correction to $e^{+}e^{-} \leftrightarrow \nu_{L}\Bar{\nu}_{L}$ interactions along with finite temperature QED corrections to the electromagnetic plasma density and effect of neutrino oscillations.} Future CMB experiment CMB Stage IV (CMB-S4) is expected reach a much better sensitivity of $\Delta {\rm N}_{\rm eff}={\rm N}_{\rm eff}-{\rm N}^{\rm SM}_{\rm eff}
= 0.06$ \cite{Abazajian:2019eic}, taking it closer to the standard model (SM) prediction. Similar precision measurements are also expected from other planned future experiments like SPT-3G \cite{Benson:2014qhw}, Simons Observatory \cite{Ade:2018sbj} as well as CMB-HD \cite{CMB-HD:2022bsz}. 

If dark radiation has sizeable interactions with the SM bath, it can be thermalised in the early universe, followed by decoupling at some stage. Depending upon the decoupling temperature and internal degrees of freedom, such dark radiation can have very specific predictions for ${\rm \Delta N_{\rm eff}}$ which can either be ruled out by existing Planck data or can be probed at future experiments. On the other hand, if DR has feeble interactions with the SM bath, it will not be thermalised but can still be produced via freeze-in. The contribution to ${\rm \Delta N_{\rm eff}}$ in such a case depends upon the DR-SM coupling. Even in the absence of any direct coupling between SM and DR, it is still possible for the latter to be produced in the early universe purely due to gravitational effects. One such possibility arises when the early universe has a sizeable abundance of primordial black holes (PBH). Depending upon the initial abundance of PBH, such DR produced solely from PBH evaporation can contribute substantially to ${\rm \Delta N_{\rm eff}}$ \cite{Hooper:2019gtx,Lunardini:2019zob}. Similar works related to production of such light beyond standard model (BSM) particles including dark radiation from PBH evaporation and phenomenological implications can be found in \cite{Baker:2022rkn, Baker:2021btk, Schiavone:2021imu, Arbey:2021ysg, Cheek:2022dbx, Cheek:2022mmy, Bhaumik:2022zdd} and references therein. In the presence of DR, entire PBH mass window in the ultra-light ballpark namely $\sim 0.1-10^8$ g can be probed in future CMB experiments while current Planck data ruling out certain PBH masses with large initial fractions.

%In this work, we study light thermalised species in presence of primordial black holes.The light thermalised species can be dirac fermion, weyl fermion, glodstone boson, massless gauge boson or graviton. %At the same time, presence of right-handed dirac neutrino can also explain small neutrino masses. Motivating from this, we consider right-handed dirac neutrino as the thermalised light species in this work. We extend the Standard Model with additional light degrees of freedom in the form of Dirac neutrinos. In the absence of additional strong interactions with the standard model, these are produced solely from the evaporation of primordial black holes, giving a contribution to $\Delta N_{\rm eff}$, as was studied in \cite{Hooper:2019gtx,Lunardini:2019zob}. There, it was shown that the constraints on ${\rm \Delta N_{\rm eff}}$ from CMB and various future experiments restrict the PBH parameter space, ruling out a significant region of initial PBH mass ($m_{\rm in}$) and initial PBH fraction ($\beta$).

Motivated by this, in the present work we consider a hybrid scenario where dark radiation can be produced both from the thermal bath as well as from PBH evaporation. While PBH can give an extra contribution to ${\rm \Delta N_{\rm eff}}$, it can also dilute any initial ${\rm \Delta N_{\rm eff}}$ generated thermally. Additionally, depending upon DR-SM interactions, PBH evaporation can lead to re-thermalisation of DR as well, putting new constraints on PBH parameters not obtained in earlier works carried out in the absence of additional DR-SM interactions. We first show the results by considering different types of dark radiation with specific decoupling temperatures and corresponding ${\rm \Delta N_{eff}}$. To illustrate the issue of re-thermalisation we consider the DR to be in the form of light Dirac neutrinos (right chiral part) although the generic conclusions reached here are valid for other types of DR as well which have sizeable interactions with the SM bath. Assuming PBH to dominate the universe at early epochs such that the constraints from ${\rm \Delta N_{\rm eff}}$ is the strongest, we also show the prospects of gravitational waves (GW) complementarity in present and future experiments. The ultra-light PBH considered in our work can lead to GW production due to PBH density fluctuations keeping it in the observable ballpark of mHz-kHz frequencies with peak amplitudes lying within reach of even LIGO-VIRGO as well as several planned experiments. We find interesting complementarity between ${\rm \Delta N_{\rm eff}}$ observations at CMB experiments and GW observations at present and near future GW detectors.

This paper is organised as follows. In section \ref{sec:sec2}, we briefly summarise the production of dark radiation from evaporating PBH. In section \ref{sec:sec3}, we consider DR production both from thermal bath and PBH evaporation with different examples of dark radiation. In section \ref{sec:sec4}, we consider Dirac active neutrino as a specific example with effective four-Fermi type interactions with the SM bath and show the possibility of re-thermalisation and its implications. We discuss the gravitational waves complementarity in section \ref{sec:sec5} and finally conclude in section \ref{sec:sec6}.

\section{Dark radiation from PBH}\label{sec:sec2}

Considering PBH to be formed in the early radiation dominated universe at a temperature, say $T_{\rm in}$, the initial mass of PBH is related to the mass enclosed in the particle horizon and is given by \cite{Fujita:2014hha, Carr:2020gox, Masina:2020xhk} 

\begin{equation}\label{eqn:Mini}
m_{\rm in}=\frac{4\pi}{3}\gamma\frac{\rho_{\rm R}(T_{\rm in})}{H^3(T_{\rm in})}, 
\end{equation}
where $\gamma\approx0.2$ \cite{Carr:2020gox}, $\rho_{\rm R}(T_{\rm in})$ is the initial radiation density and $H$ is the Hubble expansion rate. The temperature of a black hole can be related to its mass as \cite{Hawking:1974sw}
\begin{equation}\label{eq:T_BH}
T_{\rm BH}=\frac{1}{8\pi GM_{\rm BH}}\approx 1.06~\left(\frac{10^{13}\; {\rm g}}{M_{\rm BH}}\right)~{\rm GeV}.
\end{equation}
PBH may dominate the energy density of the universe depending on their initial abundance, characterized by the dimensionless parameter
\begin{equation}
\beta=\frac{\rho_{\rm BH}(T_{\rm in})}{\rho_{\rm R}(T_{\rm in})}.
\label{eqn:beta}
\end{equation}
Once PBH form\footnote{Since our primary motive is to explore the effect of PBH on dark radiation, in this work we remain agnostic about the formation mechanism of PBH. PBH with our desired mass range and initial energy density can be formed through several mechanisms, say from inflationary perturbations \cite{Hawking:1971ei, Press:1973iz, Braglia:2022phb}, Fermi-ball collapse \cite{Kawana:2021tde}, loop quantum gravity \cite{Papanikolaou:2023crz} etc.}, they lose mass through Hawking evaporation at a rate given by \cite{MacGibbon:1991tj} 
\begin{equation}
\frac{dM_{\rm BH}}{da}  = -\dfrac{\epsilon(M_{\rm BH}) \kappa}{a H} \left( \frac{1\; {\rm g}}{M_{\rm BH}} \right)^{2}. \label{eq:MBH} \end{equation}  
Here $a$ is the scale factor and $\kappa=5.34\times10^{25} \; {\rm g \; s}^{-1}$ \cite{Lunardini:2019zob}. Because of only gravitational effects, production of all particles takes place, regardless of their interaction with other particles. The evaporation function $\epsilon(M_{\rm BH})$ \cite{Lunardini:2019zob} contains contributions from both SM and BSM particles. The temperature of the thermal plasma when the PBH have completely disappeared can be found by integrating Eq. \eqref{eq:MBH}, and can be written as \cite{Bernal:2020bjf}
\begin{equation}
T_{\rm ev}\simeq\left(\frac{9g_{*}(T_{\rm BH})}{10240}\right)^{\frac{1}{4}}\left(\frac{M_{\rm P}^{5}}{m_{\rm in}^{3}}\right)^{\frac{1}{2}}\,,\label{eq:Tev}
\end{equation}
where $M_{\rm P}$ denotes the reduced Planck mass.

For the early universe to be black hole dominated, the initial energy density of PBH should satisfy \cite{Lunardini:2019zob}
\begin{align}
    \beta \geq \beta_{\text{crit}} \equiv 2.5\times 10^{-14} \gamma^{-\frac{1}{2}} \left(\frac{M_{BH}(T_{\rm in})}{10^8 \text{g}}\right)^{-1} \left(\frac{\epsilon(M_{\rm BH}(T_{\rm in}))}{15.35}\right)^{\frac{1}{2}}.\label{eq:betacr}
\end{align}
Since PBH evaporation produces all particles, including radiation that can disturb the successful predictions of BBN, we require $T_\text{ev}>T_\text{BBN}\simeq 4$ MeV. This can be translated into an upper bound on the PBH mass. On the other hand, a lower bound on PBH mass can be obtained from the CMB bound on the scale of inflation \cite{Planck:2018jri} : $H_I\equiv H(T_\text{in})\leq 2.5\times 10^{-5}\,M_P$, where $H(T_\text{in})=\frac{1}{2\,t_\text{in}}$ with $t(T_\text{in}) \propto m_\text{in}$. Using these BBN and CMB bounds together, we have a window\footnote{The range of PBH masses in this window remains typically unconstrained~\cite{Carr:2020gox}.} for allowed initial mass for ultra-light PBH that reads $0.1\,\text{g}\lesssim m_\text{in}\lesssim 4\times 10^8\,\text{g}$. We consider this allowed mass range of ultra-light PBH in the context of dark radiation. For simplicity, we consider a monochromatic mass function of PBHs implying all PBHs to have identical masses. Additionally, the PBHs are assumed to be of Schwarzschild type without any spin and charge. 

Now, if there exist any light BSM degrees of freedom or dark radiation, they can be produced directly from evaporating PBH, contributing to the effective number of relativistic degrees of freedom ${\rm N}_{\rm eff}$ defined as 
\begin{equation}
    {\rm N}_{\rm eff} = \frac{8}{7} \left ( \frac{11}{4} \right)^{4/3} \left ( \frac{\rho_{R}-\rho_{\gamma}}{\rho_{\gamma}} \right )\,,
\end{equation}
where $\rho_{R}, \rho_{\gamma}$ denote total radiation and photon densities respectively. In order to track the evolution of the energy densities, we need to  consider the following set of Boltzmann equations \cite{Lunardini:2019zob} 
\begin{eqnarray}
\frac{d{\rho}_{\rm BH}}{da} + 3 \frac{\rho_{\rm BH}}{a}& \ = \ &\frac{1}{M_{\rm BH}}\,\frac{dM_{\rm BH}}{da}\,{\rho}_{\rm BH}\,\label{eq:rhoBH},  \\
aH\frac{d{\rho}_{R}}{da}+4H \rho_{R} & \ = \ &-\frac{\epsilon_\text{R}(M_\text{\rm BH})}{\epsilon(M_\text{\rm BH})}\,\frac{aH}{M_{\rm BH}}\,\frac{dM_\text{BH}}{da}\,{\rho}_{\rm BH}\, \label{eq:rhoR}, \\
a\frac{d{\rho}_{X}^{\rm BH}}{da}+4 \rho_{X}^{\rm BH}& \ = \ & -\frac{\epsilon_{X}(M_\text{BH})}{\epsilon(M_\text{BH})}\,\frac{a}{M_{\rm BH}}\,\frac{dM_\text{\rm BH}}{da}\,{\rho}_{\rm BH}.\label{eq:RHNBH}  
\end{eqnarray}
 Here, $\rho_{X}^{\rm BH}$ denotes energy density of a light species $X$ produced solely from the evaporation of PBH.  The $\epsilon_{R}$ and $\epsilon_{X}$ are the evaporation functions of SM particles and $X$ species respectively. The combined evaporation function is denoted by $\epsilon = \epsilon_{R} + \epsilon_{X}$.  The Hubble parameter $H$ entering in the above equations is given by
\begin{equation}\label{eqn:Hubble}
H=\sqrt{\dfrac{\rho_{\rm BH}+\rho_{\rm SM}+\rho_{X}}{3M_{\rm P}^{2}}}.
\end{equation}
Since entropy is not conserved due to PBH evaporation, we track the evolution of the thermal bath separately through the equation given by 

\begin{equation}
\frac{dT}{da}  = -\frac{T}{\Delta}\left(\frac{1}{a}+\frac{\epsilon_{SM}(M_{\rm BH})}{\epsilon(M_{\rm BH})}\frac{1}{M_{\rm BH}}\frac{dM_{\rm BH}}{da}\frac{\rho_{\rm BH}}{4(\rho_{\rm SM}+\rho_{X})}\right), 
\end{equation}
where 
\begin{eqnarray}
    \Delta = 1 + \frac{T}{3 g_{* s}(T)} \frac{d g_{* s}(T)}{d T}.
\end{eqnarray}
 The extra relativistic degrees of freedom, ${\rm \Delta N_{\rm eff}}$ is defined as 
\begin{eqnarray}
    \Delta N_{\rm eff} = \frac{\rho_{X} (T_{\rm eq})}{\rho_{\nu_{L}, 1}(T_{\rm eq})}.
\end{eqnarray}
Here $\rho_{\nu_{L}, 1}(T_{\rm eq})$ denotes the energy density of one species of SM neutrinos at the time of matter-radiation equality. In terms of energy densities at the time of PBH evaporation, the above expression turns out to be  \cite{Lunardini:2019zob} 
\begin{align}\label{eqn:delNeff_NT}
     {\rm \Delta N_{\rm eff}^{\rm BH}} = \left\{\frac{8}{7}\left(\frac{4}{11}\right)^{-\frac{4}{3}}+{\rm N_{\rm eff}^{\rm SM}}\right\} 
    \frac{\rho_{\rm X}(T_{\rm ev})}{\rho_{\rm R}(T_{\rm ev})}
    \left(\frac{g_*(T_{\rm ev})}{g_*(T_{\rm eq})}\right)
    \left(\frac{g_{*s}(T_{\rm eq})}{g_{*s}(T_{\rm ev})}\right)^{\frac{4}{3}},
\end{align}
where ${\rm N^{\rm SM}_{\rm eff}} = 3.044$ and $\rho_{X}(T_{\rm ev}), \rho_{R}(T_{\rm ev})$ denote energy density of species $X$ and SM after PBH evaporation respectively. The relativistic degrees of freedom in energy and entropy densities are denoted by $g_*, g_{*s}$ respectively.

Now, in the case when PBH dominates the energy density of the universe at some epoch, Eq. \eqref{eqn:delNeff_NT} simplifies into 
\begin{align}\label{eqn:delNeff_NT2}
    {\rm \Delta N_{\rm eff}^{\rm BH}} = 13.714 \times
   \frac{\epsilon_{X}(M_{\rm BH})}{\epsilon(M_{\rm BH})}
    \frac{g_*(T_{\rm ev})}{g_{*s}(T_{\rm ev})^{\frac{4}{3}}}.
\end{align}
From the above equation, we can see that ${\rm \Delta N_{\rm eff}}$ remains constant for those values of PBH mass for which PBH evaporate before the electroweak scale since $g_*, g_{*s}$ remains constant at high temperatures. As we  keep on increasing the PBH mass further, ${\rm \Delta N_{\rm eff}}$ keeps on increasing. In the left panel of Fig. \ref{fig:neff_comparison}, we show the variation of ${\rm \Delta N_{\rm eff}}$ with the initial PBH mass $m_{\rm in}$, for three different types of DR which includes right chiral part of Dirac active neutrinos $\nu_R$, massless gauge boson (MGB) and Goldstone boson (GB). The current $2\sigma$ bound from Planck 2018 \cite{Planck:2018vyg} and the future sensitivity of CMB-S4 \cite{Abazajian:2019eic} are  also shown in the same figure. 

\section{Thermalised dark radiation in the presence of PBH}
%\label{sec:sec4}
%\section{Thermalised light degrees of freedom}
\label{sec:sec3}

\begin{figure}[h!]
\includegraphics[height=6.6cm,width=8.0cm,angle=0]{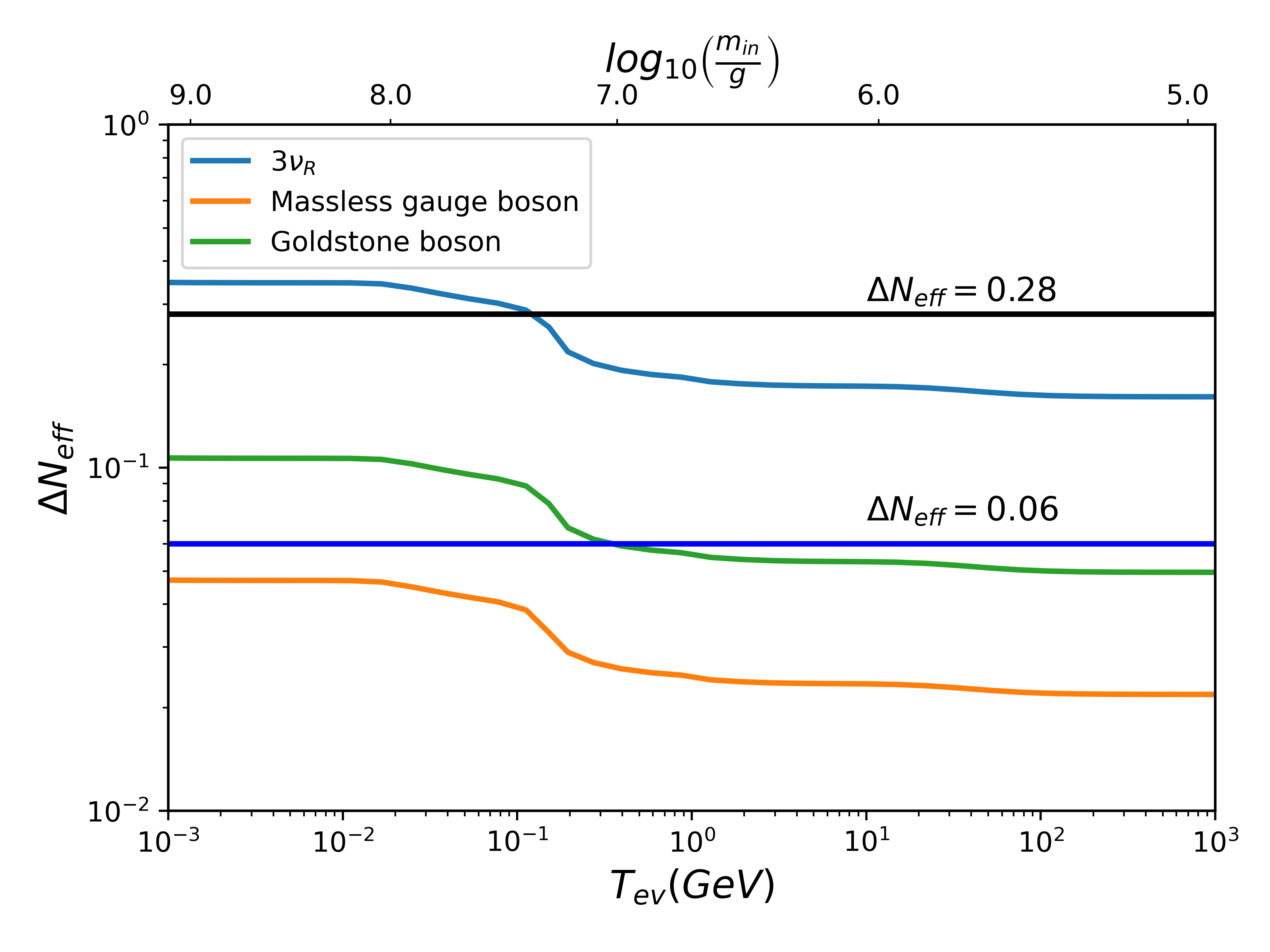}
\includegraphics[height=6cm,width=8.0cm,angle=0]{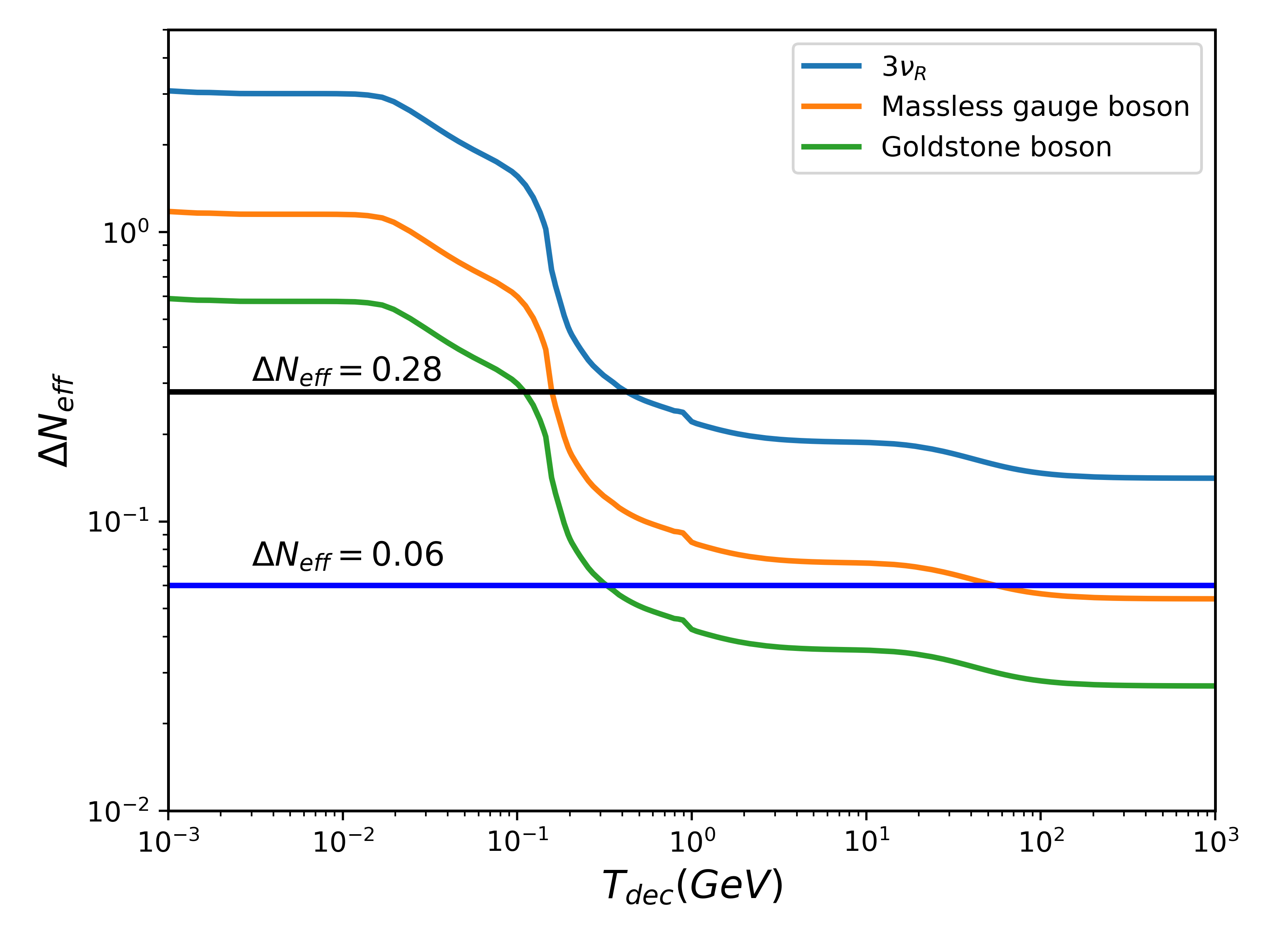}
\caption{\textit{Left panel}: Variation of non-thermal contribution to  ${\rm \Delta N_{\rm eff}}$ coming from PBH, with $T_{\rm ev}$ (or $m_{\rm in}$), for different species. \textit{Right panel}: Variation of thermal contribution to  ${\rm \Delta N_{\rm eff}}$  with $T_{\rm dec}$ for different species.}
\label{fig:neff_comparison}
\end{figure}

If we consider the extra relativistic species to have interactions with the SM bath,  ${\rm \Delta N_{\rm eff}}$ would receive a  thermal contribution which depends on the temperature at which the thermal species decouple from the bath namely $T_{\rm dec}$. It can be written as \cite{Abazajian:2019oqj} 
\begin{eqnarray}\label{eqn:delneffth}
     \Delta N_{\rm eff}^{\rm th} = 0.0267\times f n_{X} g_{X} \left(\frac{106.75}{g_{*s}(T_{\rm dec})}\right)^{4/3},
\end{eqnarray}
where $g_{X}, n_{X}$ is the internal spin degree of freedom of $X$ and number of different species of type $X$ respectively. $f$ takes the value of $7/8$ for fermion and $1$ for boson. $T_{\rm dec}$ denotes the decoupling temperature of the thermalised species. In the right panel of Fig. \ref{fig:neff_comparison}, we show the corresponding ${\rm \Delta N_{\rm eff}}$ by considering DR to be of thermal origin only. The x-axis denotes the decoupling temperature $T_{\rm dec}$ of DR from the thermal bath. As expected, a lower decoupling temperature leads to a larger contribution to ${\rm \Delta N_{\rm eff}}$.

As we will see, in the presence of PBH, the total contribution to ${\rm \Delta N_{\rm eff}}$ from thermalised DR is decided by the interplay between the thermal and the non-thermal contribution. Interestingly, the total contribution depends on when the PBH evaporates relative to the decoupling temperature of the thermalised species. Depending on the initial fractional energy density ($\beta$) and the initial mass ($m_{\rm in}$) of PBH at the time of formation, we can broadly divide our study into two categories: (A) PBH dominates the energy density of the universe at some epoch, and (B) PBH never dominates the energy density of the universe. These two possibilities lead to different observational consequences, which we discuss below.

%\section{Thermalised light species in the presence of PBH}\label{sec:sec4}

%Note that in writing the boltzmann equations in previous section, we ignore the interactions between already present thermalised RH neutrinos ($\nu_{R}^{T}$) and neutrinos emitted from PBH ($\nu_{R}^{BH}$). This can be done because, if the thermalised $\nu_{R}^{T}$ decouple before PBH evaporation, the interactions between $\nu_{R}^{T}$ and $\nu_{R}^{BH}$ will have no effect. On the other hand, if $\nu_{R}^{T}$ neutrinos decouple after PBH evaporation, all the RH neutrinos emitted from PBH will get thermalised with the  $\nu_{R}^{T}$. If the $\nu_{R}^{T}$ neutrino decouple just after PBH evaporation, the interaction between  PBH emmitted neutrino and $\nu_{R}^{T}$ will have some effect. However, with the assumption that thermalised RH neutrino decouple instantaneously, PBH evaporate instantaneously and that the emitted neutrinos from PBH get thermalised instantaneously, we can safely neglect the interactions and proceed further. 

%\section{Contribution to ${\rm \Delta N_{\rm eff}}$ : Analytical approximation}

\subsection{PBH domination}
The evaporation of PBH will produce all the SM particles along with DR. In the hybrid scenario considered here, DR can be produced gravitationally from PBH as well as from the SM bath by virtue of non-standard DR-SM interactions. Therefore, we can think of the total contribution to the extra number of relativistic species, ${\rm \Delta N_{\rm eff}}$ as the contribution from DR emitted from PBH and the already existing thermal counterpart, which was thermalised followed by decoupling from the thermal bath.

The decoupling temperature of extra light species, $X$ depends on its coupling with SM particles. A smaller value of coupling can lead to an early decoupling. In such a scenario, the duration between decoupling and PBH evaporation can be quite long. Thus, to begin with, we neglect the interaction between the decoupled $X$ and $X$ produced from PBH. So, both decoupled and light species from PBH can be treated as different species while considering their individual contribution to ${\rm \Delta N_{\rm eff}}$. In a follow-up section, we explicitly consider a specific type of DR namely Dirac  type active neutrinos with four-Fermi type interactions of right chiral part with the SM and study the validity of this assumption. On the other hand, for a larger value of DR-SM coupling, the PBH evaporation can occur much before the decoupling of light species, $X$. In that case, the emitted $X$ from PBH will be thermalised completely with the SM bath, and both emitted and preexisting thermal $X$ will behave as a single species.

\begin{figure}[h!]
\includegraphics[height=8cm,width=8.0cm,angle=0]{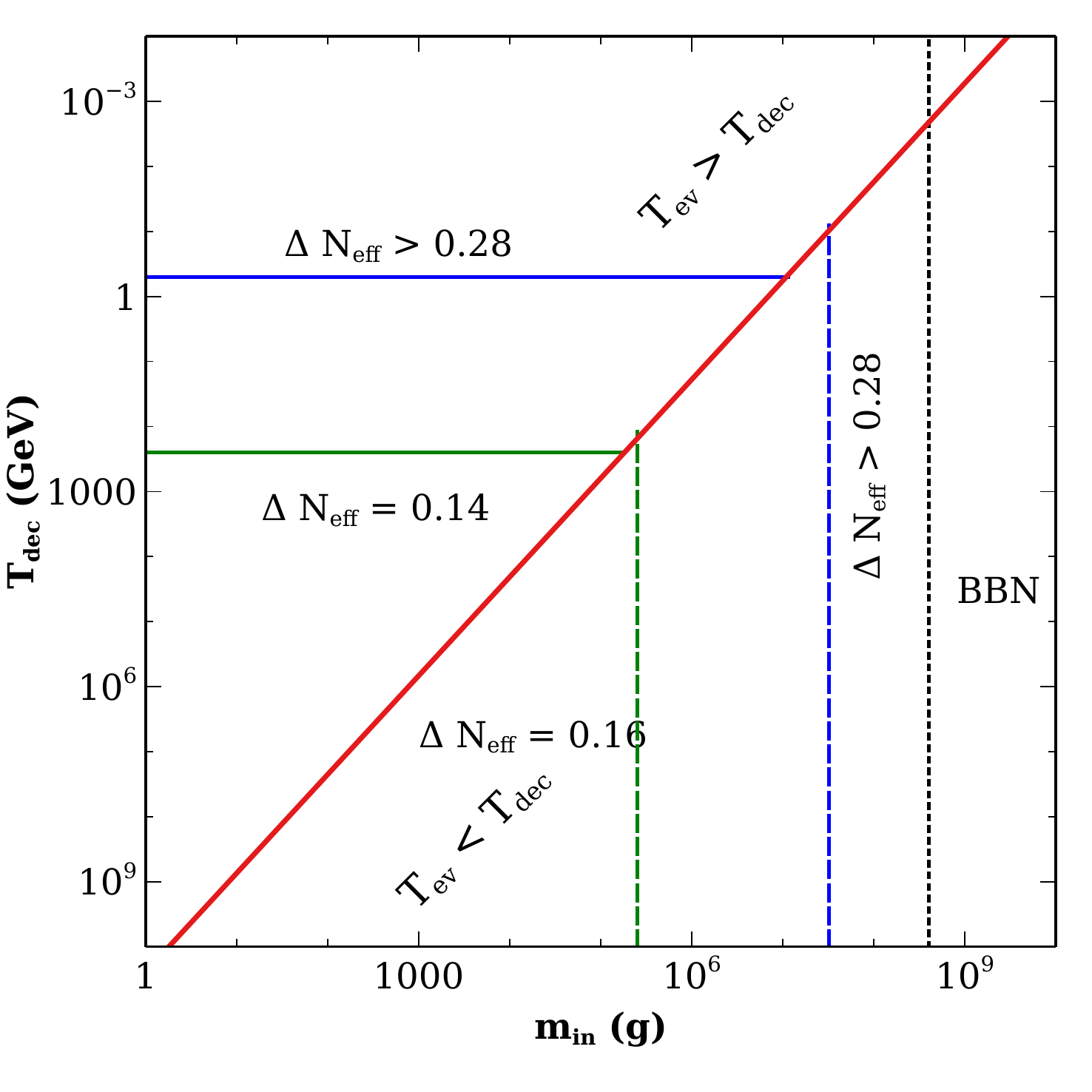}
\includegraphics[height=8cm,width=8.0cm,angle=0]{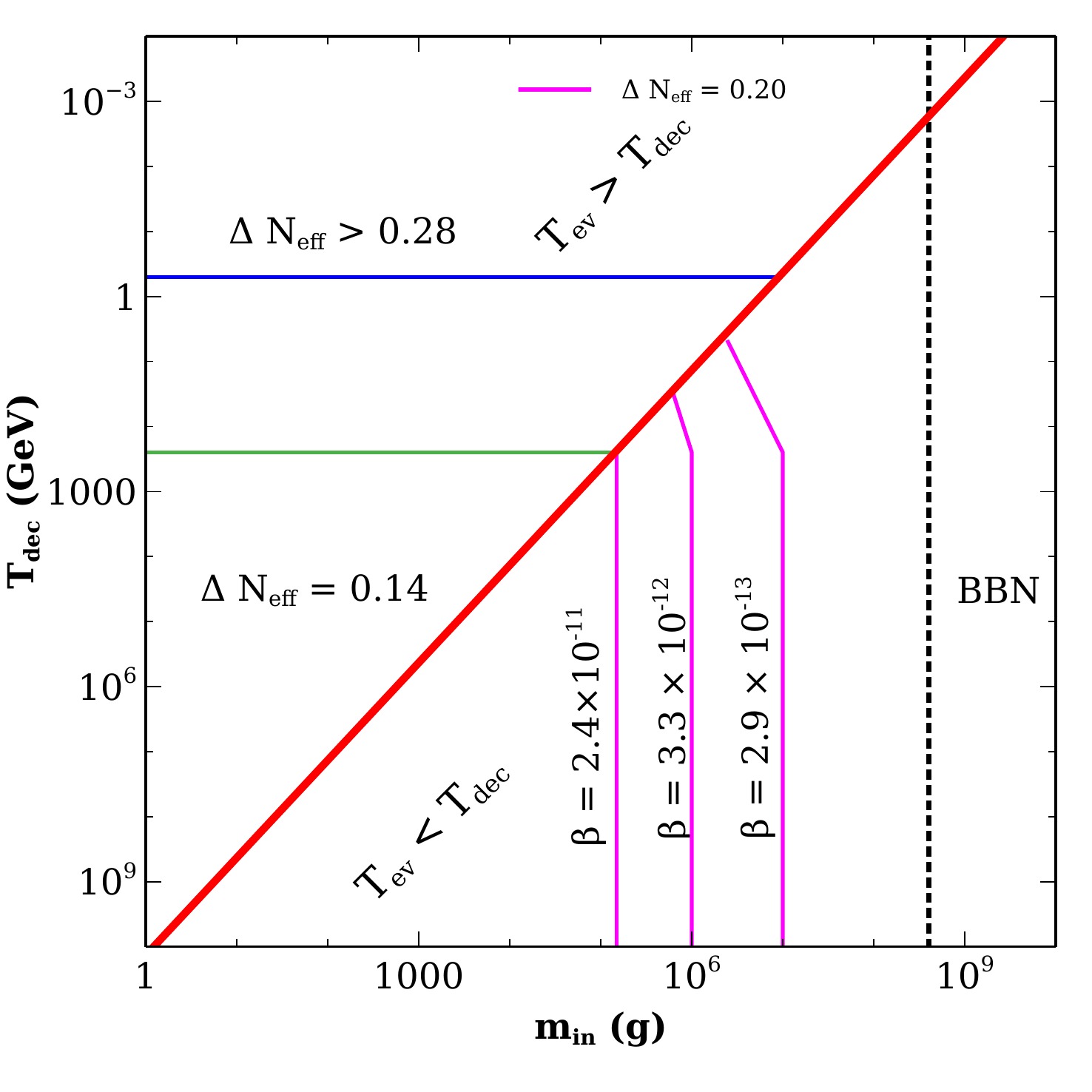}
\caption{${\rm \Delta N_{\rm eff}}$ contours in the plane of initial PBH mass versus decoupling temperature. Here, we consider the extra light species $X$ to be $\nu_R$. The left plot is for the case where PBH dominate the energy density at the time of evaporation, whereas in the right panel, PBH energy density is always less than that of radiation. The black vertical dashed line indicates the upper bound on PBH mass from BBN limit.}
\label{fig:Mass_vs_tdec}
\end{figure}

Here, we make an estimate of ${\rm \Delta N_{\rm eff}}$ in the two limits discussed above. The evaporation temperature of PBH is connected to the initial PBH mass $m_{\rm in}$ through Eq. \eqref{eq:Tev}. Assuming that the species $X$ decouples at the same temperature as PBH evaporation, we can draw the solid red line in Fig. \ref{fig:Mass_vs_tdec} corresponding to $T_{\rm ev} = T_{\rm dec}$. Now, depending upon the values of its coupling with SM and $m_{\rm in}$, we can have the two different scenarios, i.e. either $T_{\rm ev} > T_{\rm dec}$ (above the red line)  or  $T_{\rm ev} < T_{\rm dec}$ (below the red line). We discuss the dynamics for these two cases below. 

\subsubsection{PBH evaporation before decoupling of thermalised DR} 
\label{subsub1}
We first consider the scenario, where the PBH evaporates  earlier than the decoupling of thermalised species $X$.  In such a case, the species $X$ produced from PBH evaporation will also get thermalised. Therefore, the contribution to ${\rm \Delta N_{\rm eff}}$ comes only from the thermalised $X$, and is given by Eq. \eqref{eqn:delneffth}. For example, if $X$ is considered to be the right chiral parts of Dirac active neutrinos, this equation reduces to
\begin{eqnarray}
    {\rm \Delta N_{\rm eff}} \simeq 0.027\times2\times3\times\frac{7}{8} \left(\frac{106.75}{g_{*}(T_{\rm dec})}\right)^{4/3} \simeq 0.14175\times \left(\frac{106.75}{g_{*}(T_{\rm dec})}\right)^{4/3}.
\end{eqnarray}
Thus, ${\rm \Delta N_{\rm eff}}$ becomes independent of $m_{\rm in }$. For decoupling temperature $T_{\rm dec}$  $\lesssim 500$ MeV, ${\rm \Delta N_{\rm eff}}$ is more than $0.28$, the current Planck bound at $2\sigma$ level. This is indicated by the region above the solid blue line in Fig. \ref{fig:Mass_vs_tdec}. A larger decoupling temperature reduces the contribution to ${\rm N_{\rm eff}}$. Note that this also requires a smaller value of PBH mass to make sure that PBH evaporates earlier compared to the decoupling epoch. Above a certain value of $T_{\rm dec}$, the contribution to ${\rm \Delta N_{\rm eff}}$ saturates at a value of $0.14$, indicated by the triangular region between solid green and solid red lines of Fig. \ref{fig:Mass_vs_tdec}.

\subsubsection{PBH evaporation after decoupling of thermalised DR}
\label{subsub2}
Next, we consider the scenario where PBH evaporates after the thermal decoupling of species $X$. In this case, ${\rm \Delta N_{\rm eff}}$ can, in principle, get contributions from both sources. We can write 
\begin{align} \label{eq:PBH_dom_Neff}
    {\rm \Delta N_{\rm eff}} = {\rm \Delta N_{\rm eff}^{BH} }+ {\rm \Delta N_{\rm eff}^{th}}\,,
\end{align}
where the first term on the right hand side (RHS) denotes the contribution from PBH and the second term indicates the thermal contribution. Now, for PBH domination, evaporation of PBH will lead to entropy injection to the thermal bath. Thus the thermal contribution from $X$ to ${\rm \Delta N_{\rm eff}}$ gets diluted. Taking into account the effect of dilution, the above equation for Dirac active neutrinos can be written as
\begin{eqnarray} \label{eq:PBH_dom_Neff_1}
    {\rm \Delta N_{\rm eff}} = 0.772 \frac{g_{*}(T_{\rm ev})}{(g_{*s}(T_{\rm ev}))^{4/3}} + 0.14175\times \left(\frac{106.75}{g_{*}(T_{\rm dec})}\right)^{4/3} \xi^{-4/3},
\end{eqnarray}
where the parameter $\xi$ quantifies decrease in ${\rm \Delta N_{\rm eff}}$ due to entropy dilution. It can be written as the ratio of the comoving entropy density at PBH evaporation $(T_{\rm ev})$ to that at thermal decoupling of X $(T_{\rm dec})$
\begin{eqnarray} \label{ent_dilution}
    \xi = \frac{S_{\rm ev}}{S_{\rm dec}} = \frac{g_{*s} (T_{\rm ev})a^{3} (T_{\rm ev}) T^{3}_{\rm ev}}{g_{*s} (T_{\rm dec})a^{3} (T_{\rm dec})T^{3}_{\rm dec}}.
\end{eqnarray}
Without any entropy injection, conservation of entropy gives $\xi=1$. For entropy injection due to PBH evaporation, we have $\xi>1$. It turns out that for PBH dominated region, $\xi \gg 1$, diluting the contribution from thermal $\nu_R$ or X in general. Thus, one can safely neglect the second term in the RHS of Eq.  \eqref{eq:PBH_dom_Neff}. Hence, only the X from PBH will contribute to ${\rm \Delta N_{\rm eff}}$. PBH mass greater than around $4\times10^{7}$ g leads to ${\rm \Delta N_{\rm eff}}$ greater than the Planck $2\sigma$ bound, as shown earlier (cf. left panel of Fig. \ref{fig:neff_comparison}).

\subsection{Radiation domination}

If the ratio of the initial energy density of PBH to that of radiation energy density is less than $\beta_{\rm crit}$, the early universe remains dominated by radiation only. Similar to the PBH-dominated universe, we can also have the same two scenarios, namely $T_{\rm ev} > T_{\rm dec}$ and $T_{\rm ev} < T_{\rm dec}$. The solid red line in the right panel plot of Fig. \ref{fig:Mass_vs_tdec} corresponds to $T_{\rm dec} = T_{\rm ev}$. We discuss the two scenarios corresponding to either side of this solid red line below.

%This line remains more or less same from the PBH dominated case, as the PBH evaporation temperature while radiation dominates is only $2/\sqrt{3}$ times the evaporation temperature while PBH dominates.

\subsubsection{PBH evaporation before decoupling of thermalised DR}
Similar to the case of PBH domination, here  the thermal contribution is the only contributing factor to ${\rm \Delta N_{\rm eff}}$. Hence the region and bounds are same as that of PBH domination, as already discussed in \ref{subsub1}.

\subsubsection{PBH evaporation after decoupling of thermalised DR}
Here, the total contribution to ${\rm \Delta N_{\rm eff}}$ can be written in the form of Eq. \eqref{eq:PBH_dom_Neff}, with the non-thermal contribution ${\rm \Delta N_{\rm eff}^{\rm BH}}$ given by Eq. \eqref{eqn:delNeff_NT}. ${\rm \Delta N_{\rm eff}^{\rm BH}}$ is evaluated by solving the Boltzmann equations \eqref{eq:rhoBH}-\eqref{eq:RHNBH}. Unlike the case of PBH domination, here the contribution from the thermalised DR can not be neglected as the entropy injection is negligible (i.e. $\xi \sim 1$) for $\beta < \beta_{\rm crit}$. Hence, ${\rm \Delta N_{\rm eff}}$ will bear contribution from both thermalised as well as PBH generated DR. Also, for PBH domination, ${\rm \Delta N_{\rm eff}^{\rm BH}}$ is independent of the parameter $\beta$, whereas in the present scenario $\beta$ plays a significant role. The role of $\beta$ can be seen from the magenta coloured lines shown in the right panel plot of Fig. \ref{fig:Mass_vs_tdec}. The magenta coloured lines denote a total ${\rm \Delta N_{\rm eff} = 0.20}$. As the initial mass of PBH increases, its contribution to ${\rm \Delta N_{\rm eff}}$ also increases. As a result, one needs to reduce the value of $\beta$ to get same ${\rm \Delta N_{\rm eff}}$. Let us consider the magenta line with $m_{\rm in}=10^{7}$ g. The straight vertical portion of the line indicates a thermal contribution of $0.14$ and a PBH contribution of $0.06$. The contribution from PBH can be set to the required value by adjusting $\beta$. Decreasing $T_{\rm dec}$ in the right panel plot up to $\sim 200$ GeV leads to no changes in the thermal contribution as $g_{*s}$ remain constant, giving a constant contribution to ${\rm \Delta N_{\rm eff}^{\rm th}}$. Decreasing $T_{\rm dec}$ below $\sim 200$ GeV, the thermal contribution increases due to the reasons discussed earlier. So, one requires a smaller contribution of ${\rm \Delta N_{\rm eff}^{\rm BH}}$ in order to maintain a total contribution of 0.20. This can be done by either decreasing PBH mass or by decreasing $\beta$. Here we decrease the PBH mass keeping $\beta$ constant. This leads to the bending of magenta lines for  $T_{\rm dec} < 200$ GeV.

\subsection{Summary of results}

\begin{figure}
    \centering
    \includegraphics[scale=0.6]{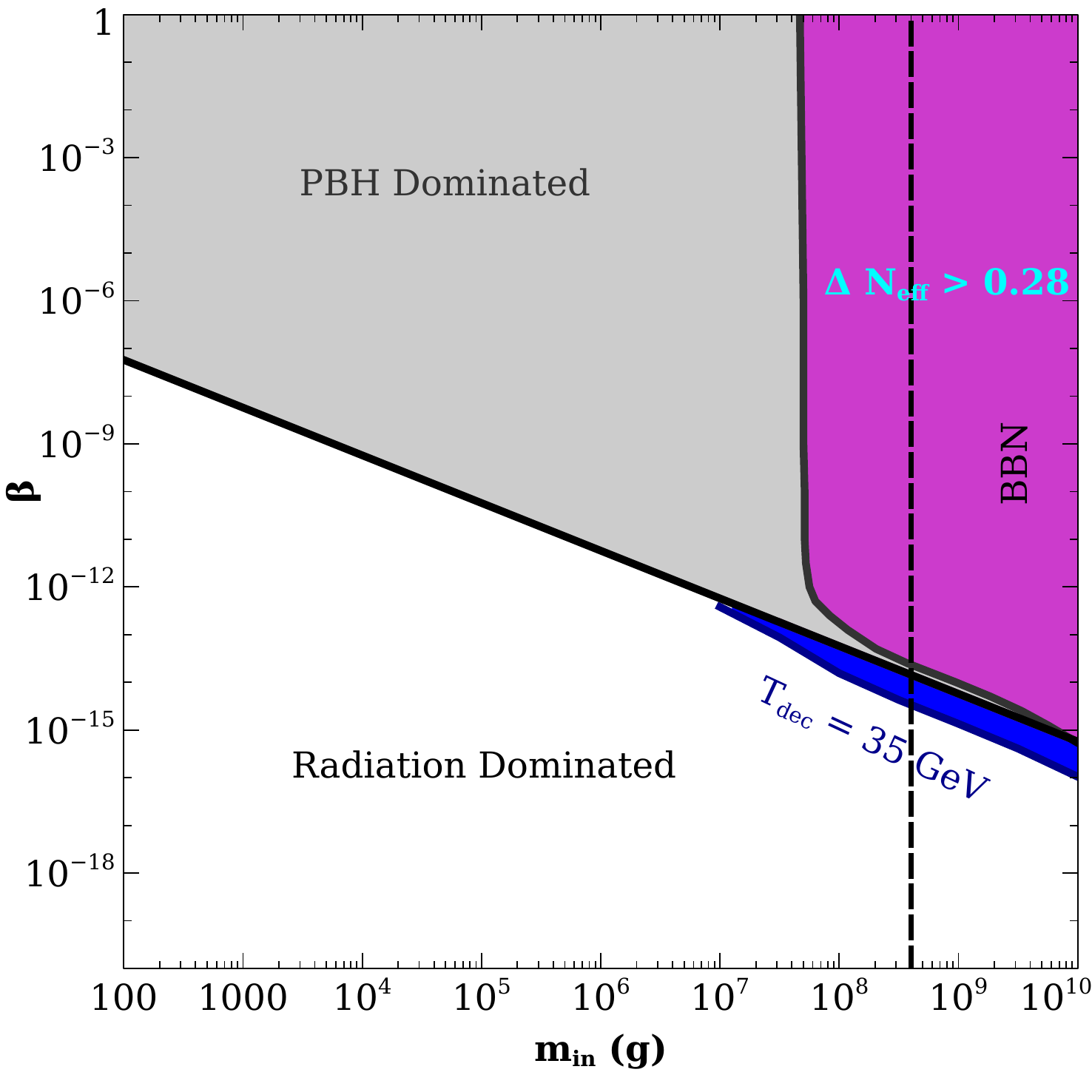}
    \caption{ ${\rm \Delta N_{\rm eff}}$ contours in the $\beta$-$m_{\rm in}$ plane. The magenta and blue region are the portion for PBH and radiation domination respectively where ${\rm \Delta N_{\rm eff}}>0.28$. The black vertical dashed line indicates the upper bound on PBH mass from BBN limit.}
    \label{fig:beta_vs_M}
\end{figure}

Now let us look at the combined results of PBH and radiation domination in the $\beta$-$m_{\rm in}$ plane. In the PBH dominated case, the magenta-shaded region in Fig. \ref{fig:beta_vs_M} shows the portion of parameter space where ${\rm \Delta N_{\rm eff}}>0.28$, corresponding to PBH initial mass $m_{\rm in} \gtrsim 4\times10^{7}$g (cf. blue-dashed contour of left panel plot in Fig. \ref{fig:Mass_vs_tdec}). Now, this constraint is only valid if $\nu_R$ (or DR, in general) decouples from the bath before PBH evaporation i.e. if  $T_{\rm dec} > 150$ MeV (intersection of the blue dashed contour with the solid red line in left panel plot of Fig. \ref{fig:Mass_vs_tdec}). If $\nu_R$ decouples after PBH evaporation, then ${\rm \Delta N_{\rm eff}}$ will bear contribution from thermalised $\nu_R$ (existing thermal $\nu_R$ + $\nu_R$ from PBH that will also get thermalised). In such a case, no constraint on the PBH parameters $\beta$ or $m_{\rm in}$ can be obtained, since the contribution is purely thermal.

In the radiation-dominated region, if the thermalised $\nu_R$ decouples earlier, then even for very small $\beta$, one expects minimum contribution of ${\rm \Delta N^{th}_{eff}}= 0.14$ to ${\rm \Delta N_{\rm eff}}$. For example, let us fix $T_{\rm dec}\approx 35$ GeV. So, for PBH mass greater than $\sim 7\times10^{5}$ g, the decoupling of $\nu_R$ occurs before PBH evaporation. The thermal contribution to ${\rm \Delta N_{\rm eff}} $ is about $ \approx 0.17$. So, if the PBH contribution to ${\rm \Delta N_{\rm eff}}$ is more than $0.11$, the total ${\rm \Delta N_{\rm eff}}$ would exceed 0.28. The blue shaded region in Fig. \ref{fig:beta_vs_M} denotes the area for which ${\rm \Delta N_{\rm eff}} > 0.28$. Note that there is a discontinuity between the magenta and blue shaded regions. This is because, we have assumed that in the PBH dominated era, the thermal contribution to ${\rm \Delta N_{\rm eff}}$ will be totally diluted away due to subsequent PBH evaporation and hence can be neglected. However, in the vicinity of the solid red line, i.e. $\beta \sim \beta_{\rm crit}$, the dilution would be smaller leading to a non-negligible thermal contribution  which should be counted for. At the same time, for the radiation-dominated case, we have assumed no entropy dilution for the thermal contribution, which is not completely true in the vicinity of $\beta \sim \beta_{\rm crit}$.  This is due to the fact that even a sub-dominant but sizeable abundance of PBH can lead to some entropy dilution, howsoever small. All these intricacies will be taken care of when we explicitly solve the relevant Boltzmann equations by properly taking into account the entropy dilution factor near the transition region from PBH to radiation domination, which we discuss below.

%\subsection{Numerical Results}

We perform a numerical analysis to evaluate the total contribution to  ${\rm \Delta N_{\rm eff}}$. For this purpose, we use the publicly available code FRISBHEE \cite{Cheek:2022dbx} to calculate the non-thermal contribution ${\rm \Delta N_{\rm eff}^{\rm BH}}$, and to properly include the entropy dilution factor (cf. Eq. \eqref{ent_dilution}), which goes into the calculation of the thermal contribution ${\rm \Delta N_{\rm eff}^{\rm th}}$.  For illustrative purposes, we consider the DR or the extra light species $X$ to be of three types namely, (i) $\nu_R$, (ii) Goldstone boson and (iii) massless gauge boson.

\subsubsection{Dirac neutrino}

\begin{figure}[h!]
\includegraphics[height=6cm,width=8.0cm,angle=0]{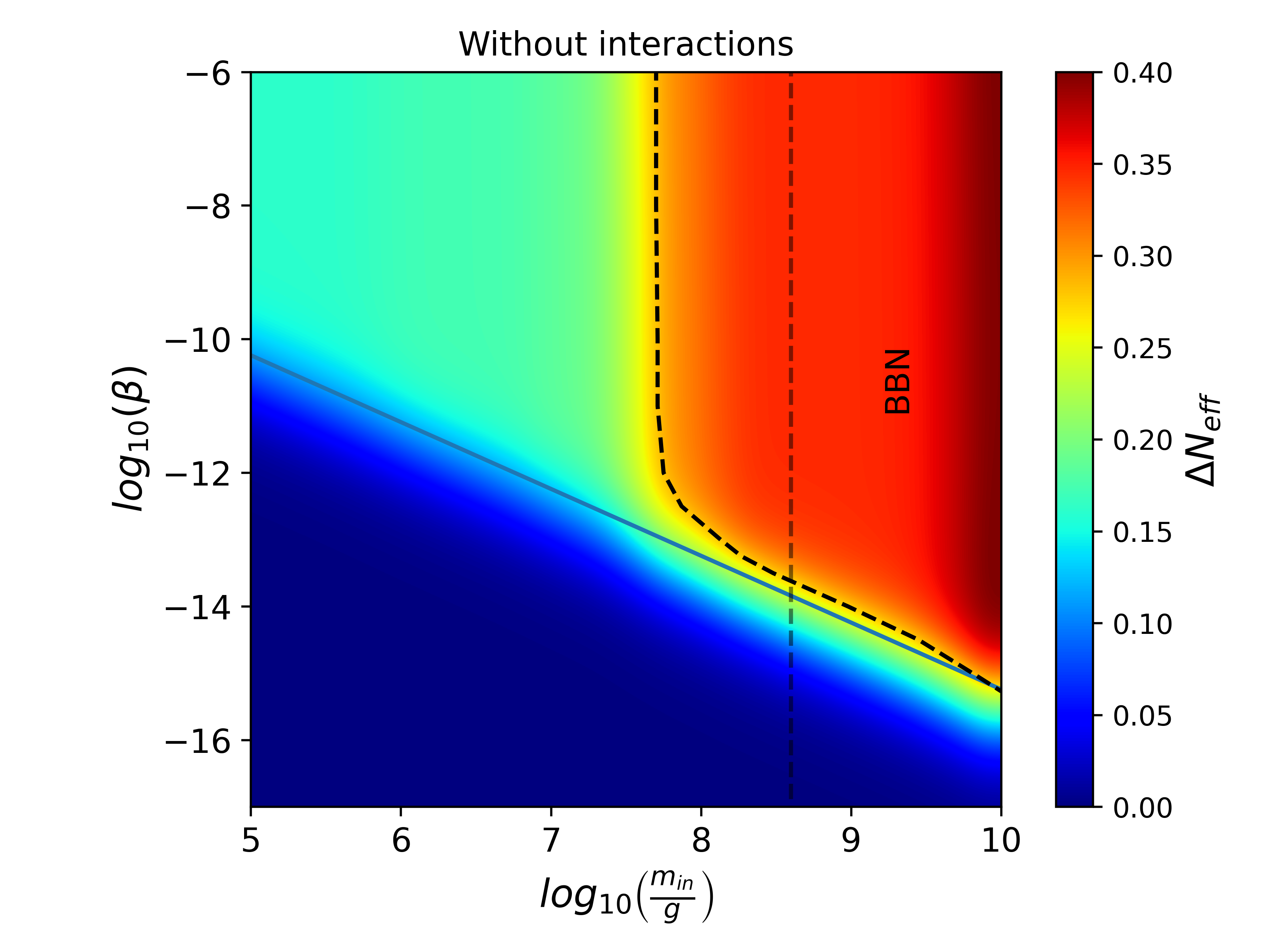}
\includegraphics[height=6cm,width=8.0cm,angle=0]{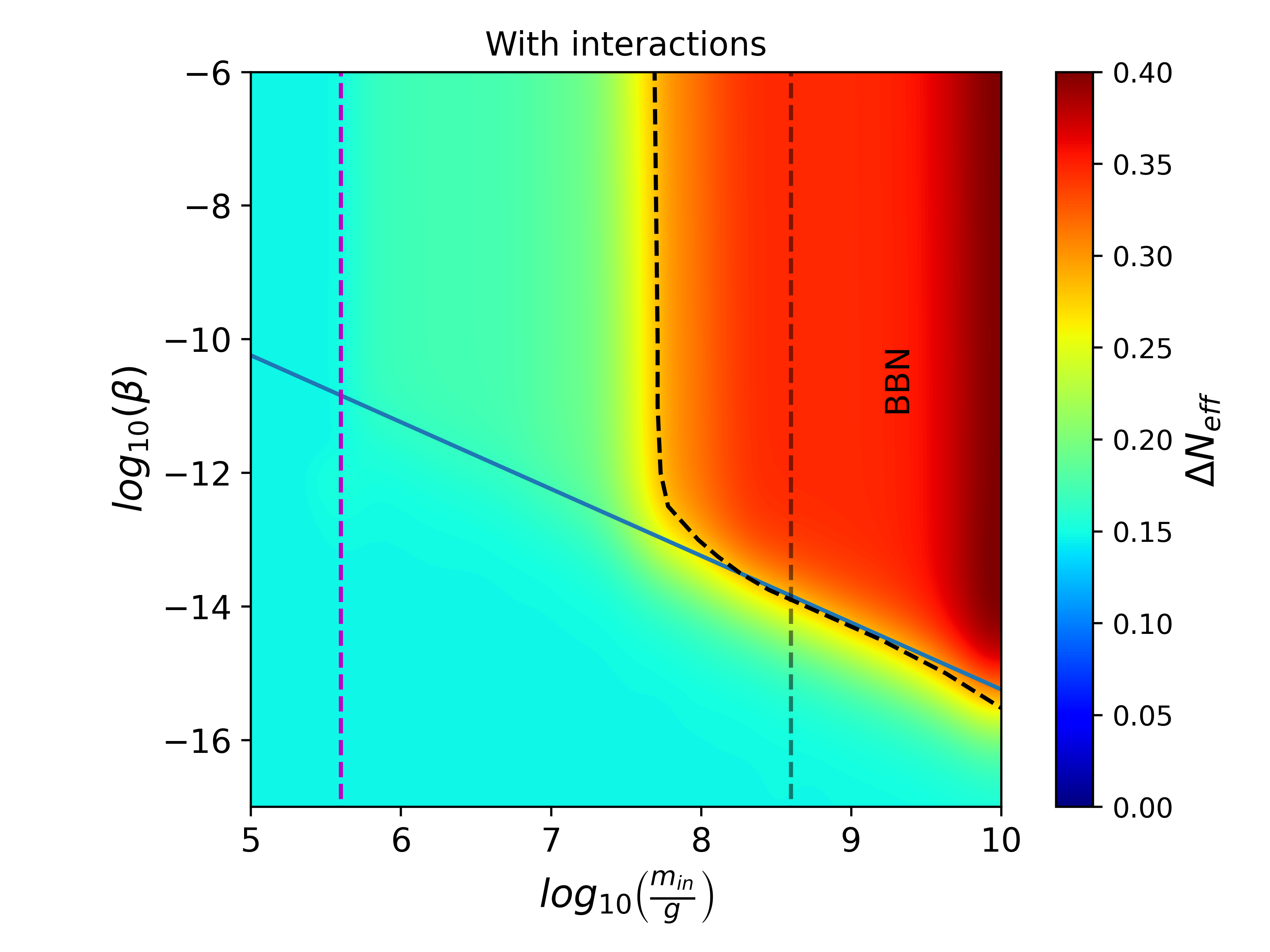}
\includegraphics[height=6cm,width=8.0cm,angle=0]{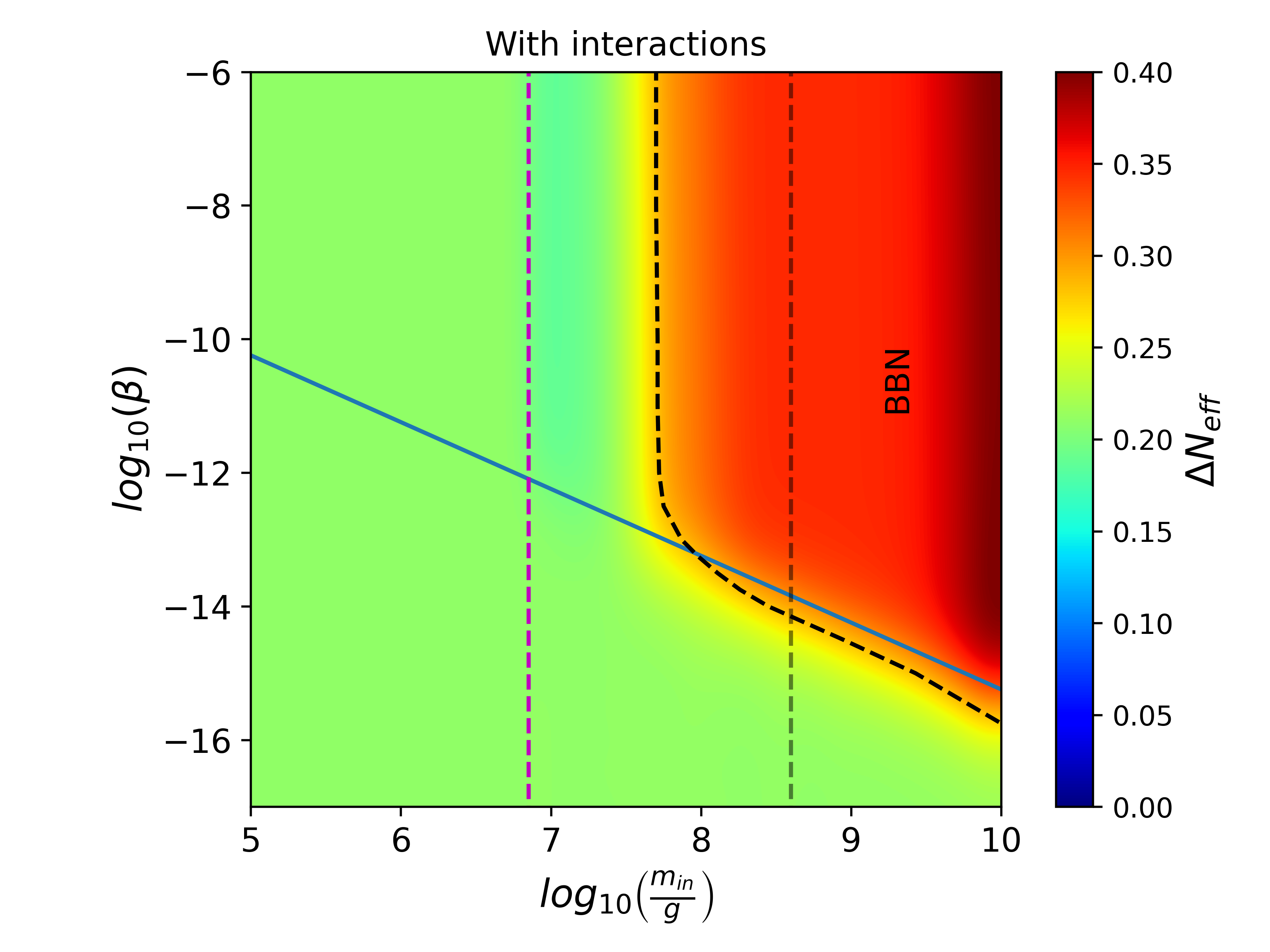}
\includegraphics[height=6cm,width=8.0cm,angle=0]{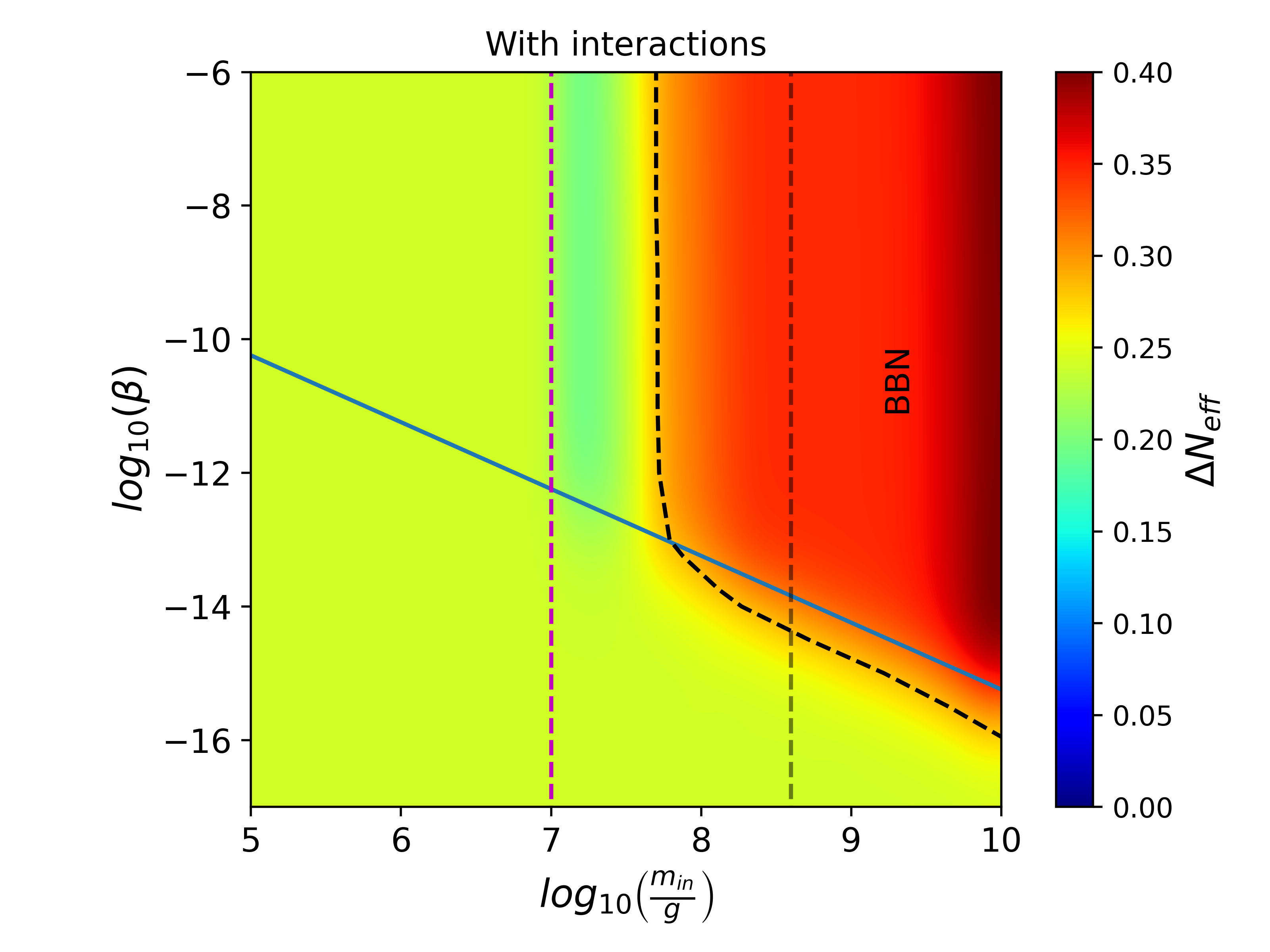}
\caption{\textit{Dirac neutrino} : Variation of ${\rm \Delta N_{\rm eff}}$ in the $m_{\rm in}$ versus $\beta$ plane for four different scenarios related to Dirac neutrinos: without any extra interactions of $\nu_R$ (top left panel), with interactions giving ${\rm \Delta N_{\rm eff}^{\rm th}} = 0.14,~\text{$T_{\rm dec} = 100$ GeV (top right panel)}, {\rm \Delta N_{\rm eff}^{\rm th}}=0.21,~\text{$T_{\rm dec} = 1$ GeV (bottom left panel)}$ and ${\rm \Delta N_{\rm eff}^{\rm th}}=0.24,~\text{$T_{\rm dec} = 850$ MeV (bottom right panel)}$ respectively. The regions right to the black dashed line indicate ${\rm \Delta N_{\rm eff}} > 0.28$ ruled out by Planck 2018 data at $2\sigma$ level. The dashed blue vertical line separates the regions $T_{\rm dec} < T_{\rm ev}$ and $T_{\rm dec} > T_{\rm ev}$ (not applicable to the top left panel due to no DR-SM interactions). The grey vertical dashed line indicates the upper bound on PBH mass from BBN limit.}
\label{fig:exact}
\end{figure}

Here, we consider $3$ species of $\nu_{R}$ as the extra light degrees of freedom, as in usual Dirac active neutrino scenarios. In Fig. \ref{fig:exact}, we show the total contribution to ${\rm \Delta N_{\rm eff}}$ in the $m_{\rm in}-\beta$ plane for four different sub-cases. In the top left panel, we show the results where $\nu_R$ is produced purely from PBH due to negligible interactions with the SM bath. The black dashed contour corresponds to ${\rm \Delta N_{\rm eff}}=0.28$, the maximum allowed by Planck 2018 data at $2\sigma$ level. This plot can be taken as a reference to compare the results in the presence of extra interactions. In the top right panel of Fig. \ref{fig:exact}, we show the results when the contribution from thermal $\nu_R$ is ${\rm \Delta N_{\rm eff}^{\rm th}}\sim 0.147$. For this particular value of ${\rm \Delta N_{\rm eff}^{\rm th}}$, the decoupling temperature should be $T_{\rm dec} \gtrsim 100$ GeV corresponding to PBH mass $m_{\rm in} \sim 4\times10^{5}$ g. PBH masses greater than $\sim 4\times 10^{5}$ g, will evaporate later satisfying $T_{\rm ev} < T_{\rm dec}$. The vertical dashed line distinguishes these two regions. For $\beta \gg \beta_{\rm crit}$, there is no change on the Planck 2018 bound compared to the results without interactions as the contribution of thermally decoupled species is negligible because of entropy dilution from PBH evaporation. However, near $\beta \sim \beta_{\rm crit}$, i.e. the transition region between PBH and radiation domination, we can see a decrease albeit small in the values of $\beta$ required to produce the same value of  ${\rm \Delta N_{\rm eff}}$ as in the case without interactions. This is because near the boundary of the transition region, the effect of the thermal contribution starts appearing, and hence we need a lower value of $\beta$ such that the non-thermal contribution ${\rm \Delta N_{\rm eff}^{\rm BH}}$ decreases, resulting in a similar value of the total ${\rm \Delta N_{\rm eff}}$. This explains slight lowering of the black-dashed contour corresponding to ${\rm \Delta N_{\rm eff}}=0.28$. Note that the minimum contribution of the thermal species is $0.147$, which is visible as we depart from the transition region. This is in sharp contrast to the results without interactions, where the value of ${\rm \Delta N_{\rm eff}}$ fades away to zero as we move deeper into the radiation domination ballpark. In the bottom panel, we have ${\rm \Delta N_{\rm eff}^{\rm th}}=$ 0.21 (left) and 0.24 (right). Due to the higher values of $ {\rm \Delta N_{\rm eff}^{\rm th}}$, and hence a lower decoupling temperature, the dashed vertical line shifts to the right corresponding to higher values of PBH mass (lower $T_{\rm ev}$). Here, the change in the transition region is more apparent, due to a decrease in the contribution from PBH namely, ${\rm \Delta N_{\rm eff}^{\rm BH}}$. We can see that for ${\rm \Delta N_{\rm eff}^{\rm th}}= 0.24$, much more parameter space near the transition region is ruled out by current Planck 2018 data compared to that in the case without interactions (cf. top left panel of Fig. \ref{fig:exact}).

\subsubsection{Goldstone boson}

\begin{figure}[h!]
\includegraphics[height=6cm,width=8.0cm,angle=0]{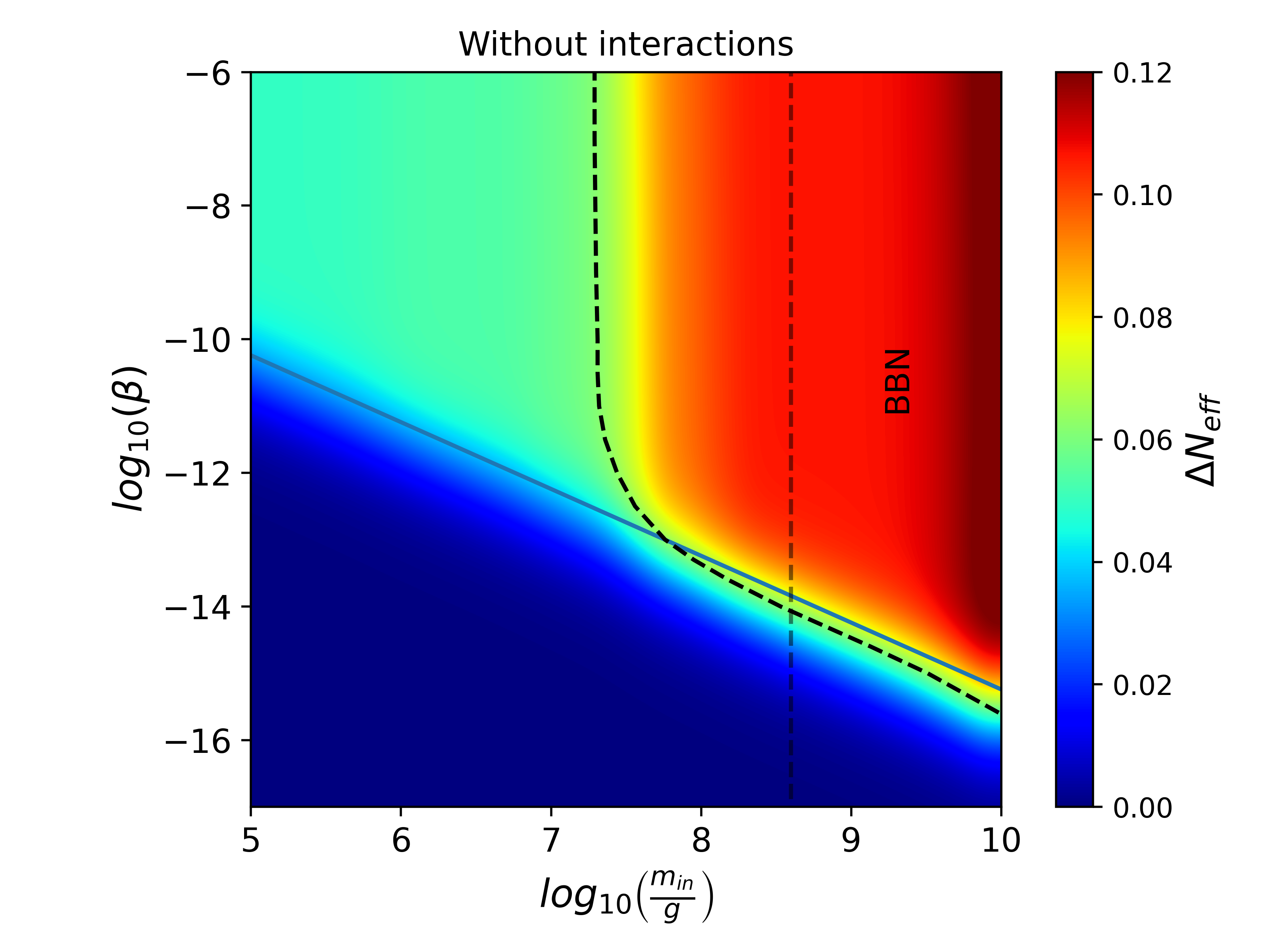}
\includegraphics[height=6cm,width=8.0cm,angle=0]{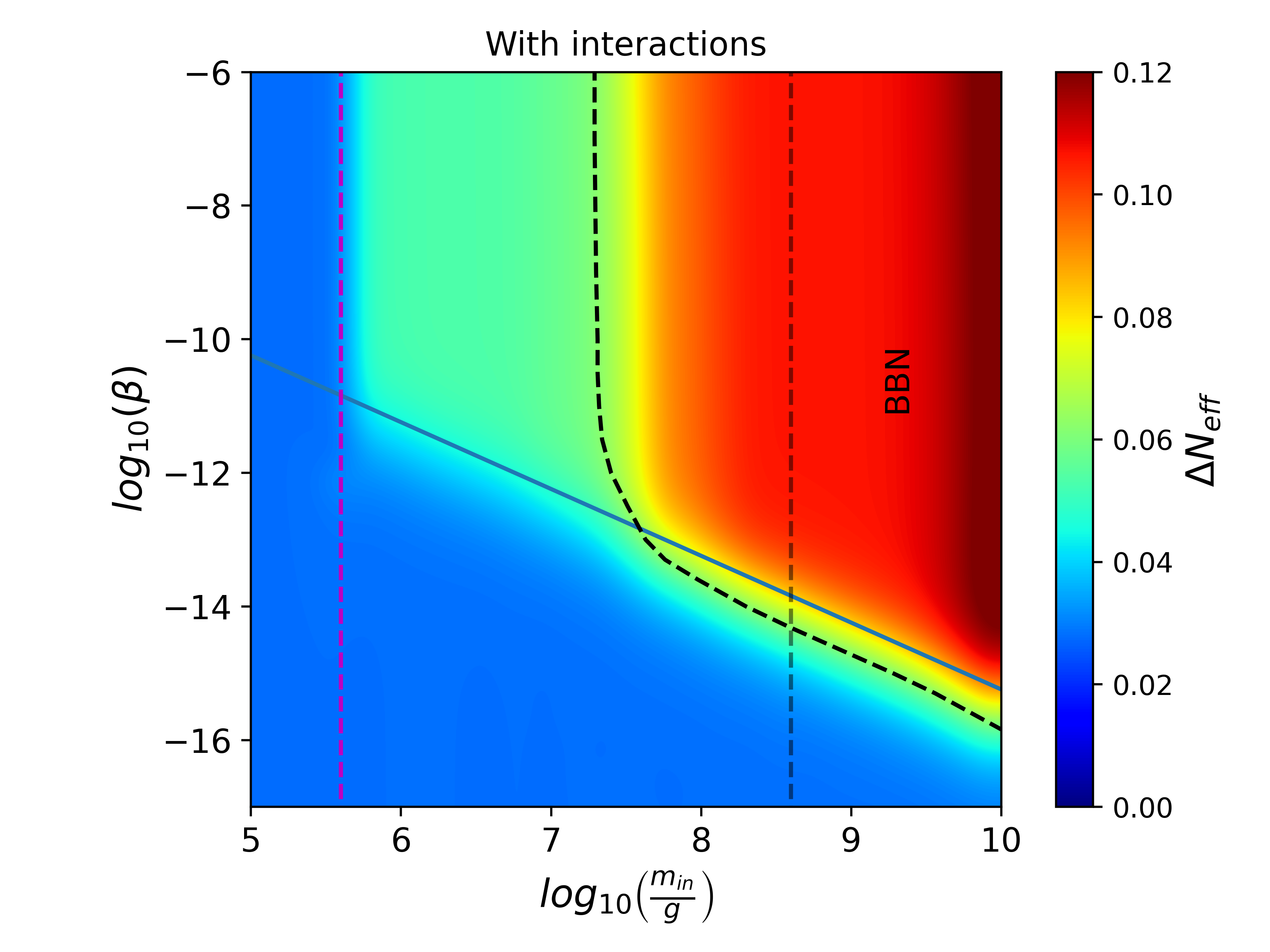}
\includegraphics[height=6cm,width=8.0cm,angle=0]{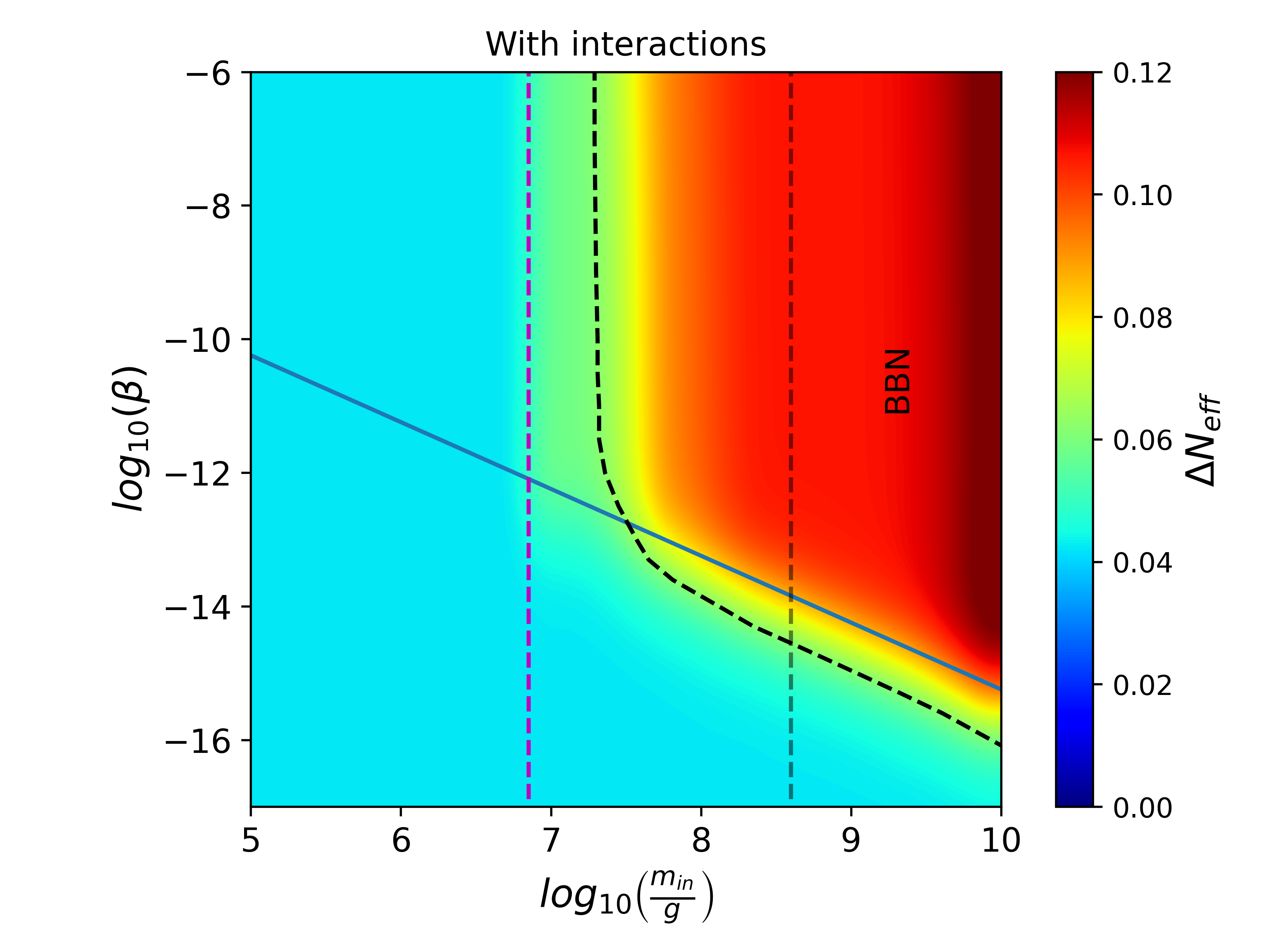}
\caption{\textit{Goldstone boson} : Variation of ${\rm \Delta N_{\rm eff}}$ in the $m_{\rm in}$ versus $\beta$ plane for three different scenarios related to Goldstone boson: without any extra interactions (top left panel), with interactions giving ${\rm \Delta N_{\rm eff}^{\rm th}} = 0.028,~\text{$T_{\rm dec} = 100$ GeV(top right panel)}$ and ${\rm \Delta N_{\rm eff}^{\rm th}}=0.042,~\text{$T_{\rm dec} = 1$ GeV (bottom panel)}$ respectively. The regions right to the black dashed contour indicate ${\rm \Delta N_{\rm eff}} > 0.06$, the CMB-S4 sensitivity. The dashed blue vertical line separates the regions $T_{\rm dec} < T_{\rm ev}$ and $T_{\rm dec} > T_{\rm ev}$. The grey vertical dashed line indicates the upper bound on PBH mass from BBN limit.}
\label{fig:exact_goldstone_boson}
\end{figure}

Let us consider the extra light species to be a Goldstone boson, a massless scalar. From Fig. \ref{fig:neff_comparison} discussed before, we can see that the maximum contribution of Goldstone boson produced solely from evaporating PBH to ${\rm \Delta N_{\rm eff}}$ is $0.10$, which is below the Planck 2018 limit. However, it comes under the sensitivity of future experiments like CMB-S4. The top left panel plot of Fig. \ref{fig:exact_goldstone_boson} shows the bound on ${\rm \Delta N_{\rm eff}}$ without any interaction. Unlike the case of $\nu_{R}$, here the region to the right of the  black dashed contour represents  ${\rm \Delta N_{\rm eff}} > 0.06$, the CMB-S4 sensitivity as Planck 2018 data do not rule out any region of the parameter space at $2\sigma$ level. The top right panel and the bottom panel plots show the corresponding bounds with interactions between GB and SM. In the top right panel plot, the decoupling temperature is $T_{\rm dec} = 100$ GeV and in the bottom panel plot, decoupling temperature is $T_{\rm dec} = 1$ GeV.  These correspond to ${\rm \Delta N^{th}_{eff}} = 0.028$ and $0.042$ respectively. Similar to the previous situation, a lower decoupling temperature gives a stronger constraint on the $\beta$ versus $m_{\rm in}$ plane. 

\subsubsection{Massless gauge boson}

 From Fig. \ref{fig:neff_comparison} discussed earlier, it can be seen that in the absence of PBH, the contribution of thermalised massless gauge boson to ${\rm \Delta N_{\rm eff}}$ is within the future CMB-S4 sensitivities for decoupling temperature $T_{\rm dec} \lesssim 100$ GeV. As a result, a wide range of interactions of MGB with SM for which $T_{\rm dec}\lesssim 100$ GeV can be probed by future CMB experiment like CMB-S4 (cf. right panel of Fig. \ref{fig:neff_comparison}). On the other hand, if PBH dominates in the early universe, the contribution to ${\rm \Delta N_{\rm eff}}$ always remains beyond the reach of future sensitivities, as shown in Fig. \ref{fig:neff_comparison} (left panel).  Thus, even if CMB-S4 rules out large coupling between MGB and SM which would produce large ${\rm \Delta N_{\rm eff}}$, such large couplings (or equivalently lower $T_{\rm dec}$) would still be allowed if we consider PBH domination in the early universe. This would be possible if we consider that PBH 
 evaporates after the decoupling of MGB and sufficiently dilutes the thermal contribution.

\section{An example: thermalised Dirac neutrino in the presence of PBH}
\label{sec:sec4}

In the previous sections, we have studied the consequence of thermalised DR in the presence of PBH while considering different examples of DR as well as PBH parameters corresponding to both PBH and radiation domination in the early universe. However, we discussed our results only in terms of decoupling temperatures of thermalised DR while being agnostic about the type of interactions with the SM bath. In this section, we consider a specific type of DR namely, light Dirac neutrinos and discuss our results for effective four-Fermi type interactions with the SM parametrised by the coupling parameter $G_{\rm eff}.$ Enhancement of ${\rm \Delta {\rm N}_{\rm eff}}$ in Dirac neutrino models (without PBH) have been studied in several recent works \cite{Abazajian:2019oqj, FileviezPerez:2019cyn, Nanda:2019nqy, Han:2020oet, Luo:2020sho, Borah:2020boy, Adshead:2020ekg, Luo:2020fdt, Mahanta:2021plx, Du:2021idh, Biswas:2021kio, Borah:2022obi, Li:2022yna, Biswas:2022fga, Biswas:2022vkq, Borah:2022enh}.

The SM extended by three right-handed singlet Dirac neutrinos with Yukawa interactions $\mathcal{L}_Y=-Y_\nu^{ab} \overline{\ell_L^a}\widetilde{\Phi}\,\nu_{b R}$, can explain tiny neutrino mass of $\mathcal{O}(\text{eV})$, with $Y_\nu\sim10^{-12}$ as a result of electroweak symmetry breaking induced by the SM Higgs doublet $\Phi$. However, such small interactions can not thermalise the $\nu_R$. Due to the smallness of Yukawa coupling, the non-thermal or freeze-in contribution to ${\rm \Delta N_{\rm eff}}$ from $\nu_R$ also remains negligible \cite{Luo:2020fdt}. Thermalisation of $\nu_R$ may be possible in the presence of non-standard interactions which we consider to be of four-Fermi type interactions with the SM neutrinos $\nu_L$ \cite{Luo:2020sho} (see appendix \ref{appendix:dirac_neutrino} for details). This allows us to work in a model-independent manner while constraining the effective interactions as well as PBH parameters. The total interaction rate of $\nu_R$ arising from such four-fermion operators can be written as 
\begin{eqnarray}\label{eq:interac}
    \Gamma &=& n^{\rm eq}_{R} \langle \sigma v \rangle \\ \nonumber
           &\simeq& \frac{3}{4} \frac{\zeta(3)}{\pi^2} 2 T^3 \frac{1}{4\pi} 4(3.151 T)^2 G_{\rm eff}^2,
\end{eqnarray}
where $G_{\text{eff}}$ is given by
\begin{eqnarray}
    G^2_{\text{eff}} = \frac{4}{3}(G_{S}-12G_{T})^2 + \frac{5}{12}(\Tilde{G}_{S} - 2 G_{V})^2\,,
\end{eqnarray}
where $G_S, \Tilde{G}_{S}, G_T, G_V$ are dimensionful couplings involved in different four-fermion operators involving $\nu_R$ and $\nu_L$ as shown in appendix \ref{appendix:dirac_neutrino}. The equilibrium number density is denoted by $n^{\rm eq}_{R}$. Comparing this interaction rate $\Gamma$ with the Hubble rate of expansion (given by Eq. \eqref{eqn:Hubble} and obtained by solving the Boltzmann equations \eqref{eq:rhoBH}-\eqref{eq:RHNBH}), we can find the decoupling temperature of $\nu_R$. Note that the Hubble parameter contains contributions from both SM radiation and PBH, which can lead to a change in the $\nu_R$ decoupling temperature compared to the standard case without PBH.

\begin{figure}[h!]
\includegraphics[height=8cm,width=8.0cm,angle=0]{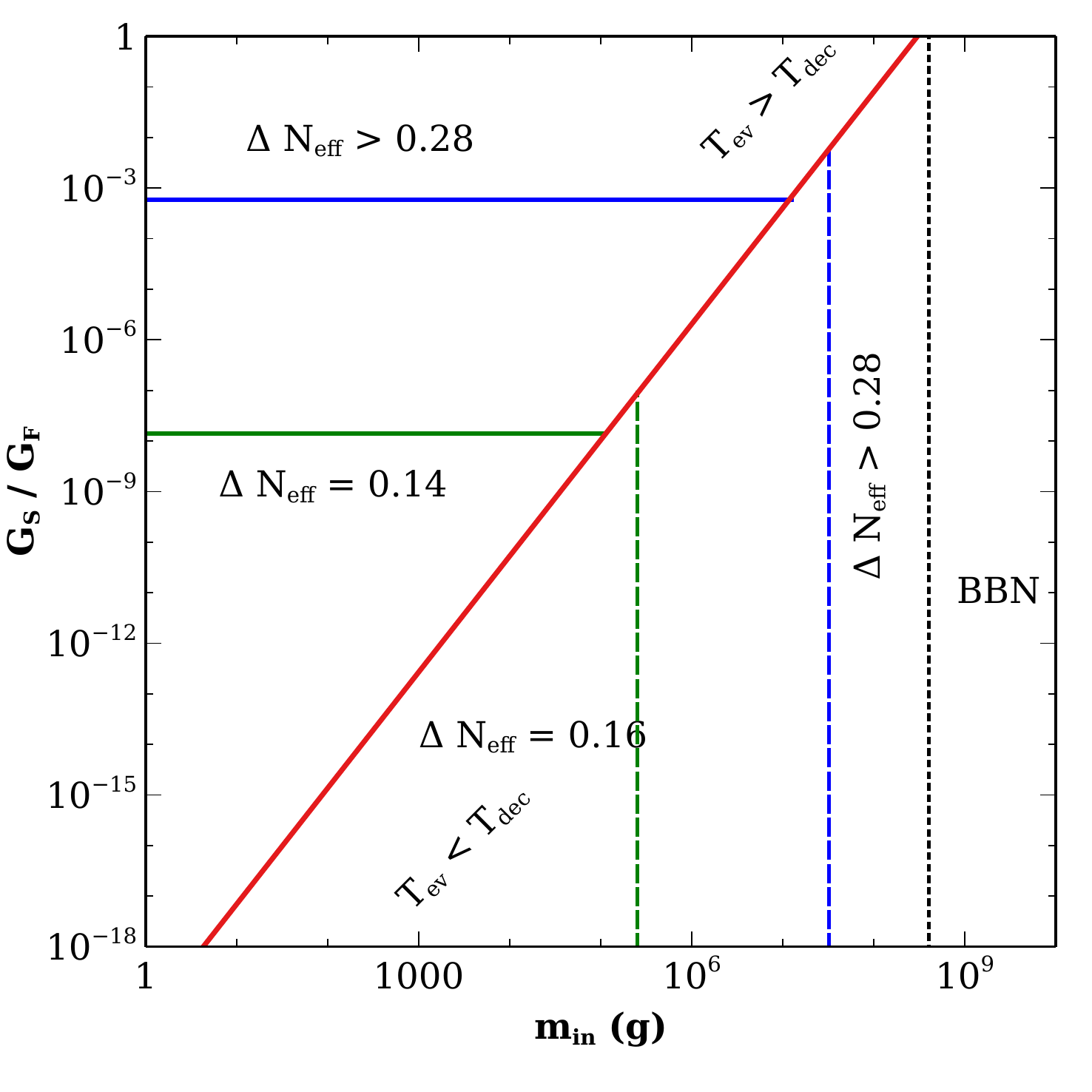}
\includegraphics[height=8cm,width=8.0cm,angle=0]{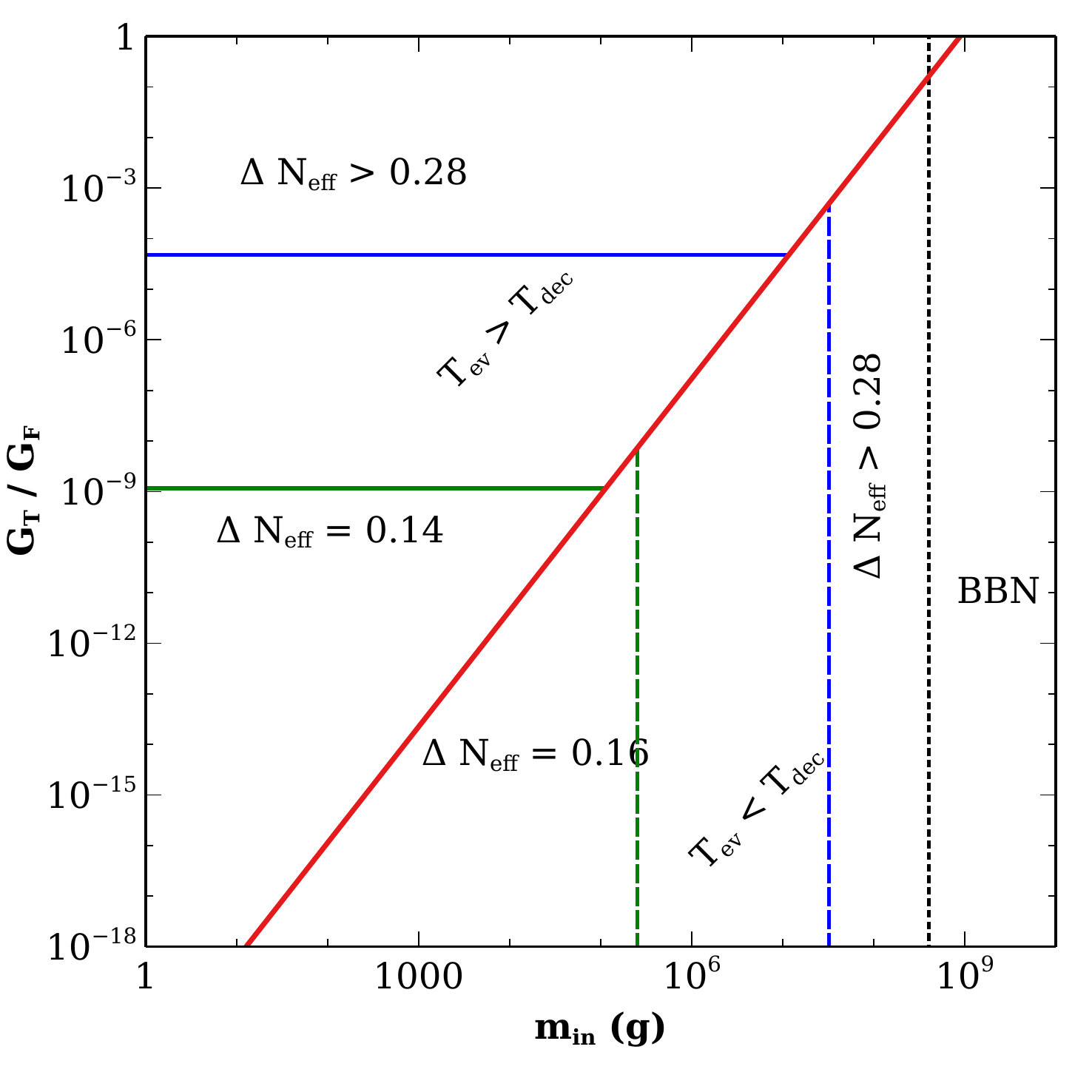}
\caption{${\rm \Delta N_{\rm eff}}$ contours in the plane of initial PBH mass versus the effective neutrino coupling for the case of PBH domination. In the left panel, $G_{T} = \Tilde{G}_{S} = G_{V} = 0$ with non-zero $G_{S}$, whereas in the right panel, we consider $G_{S} = \Tilde{G}_{S} = G_{V} = 0$ with non-zero $G_{T}$. The black vertical dashed line indicates the upper bound on PBH mass from BBN limit.}
\label{fig:PBH_dominated}
\end{figure}

\begin{figure}[h!]
\includegraphics[height=8cm,width=8.0cm,angle=0]{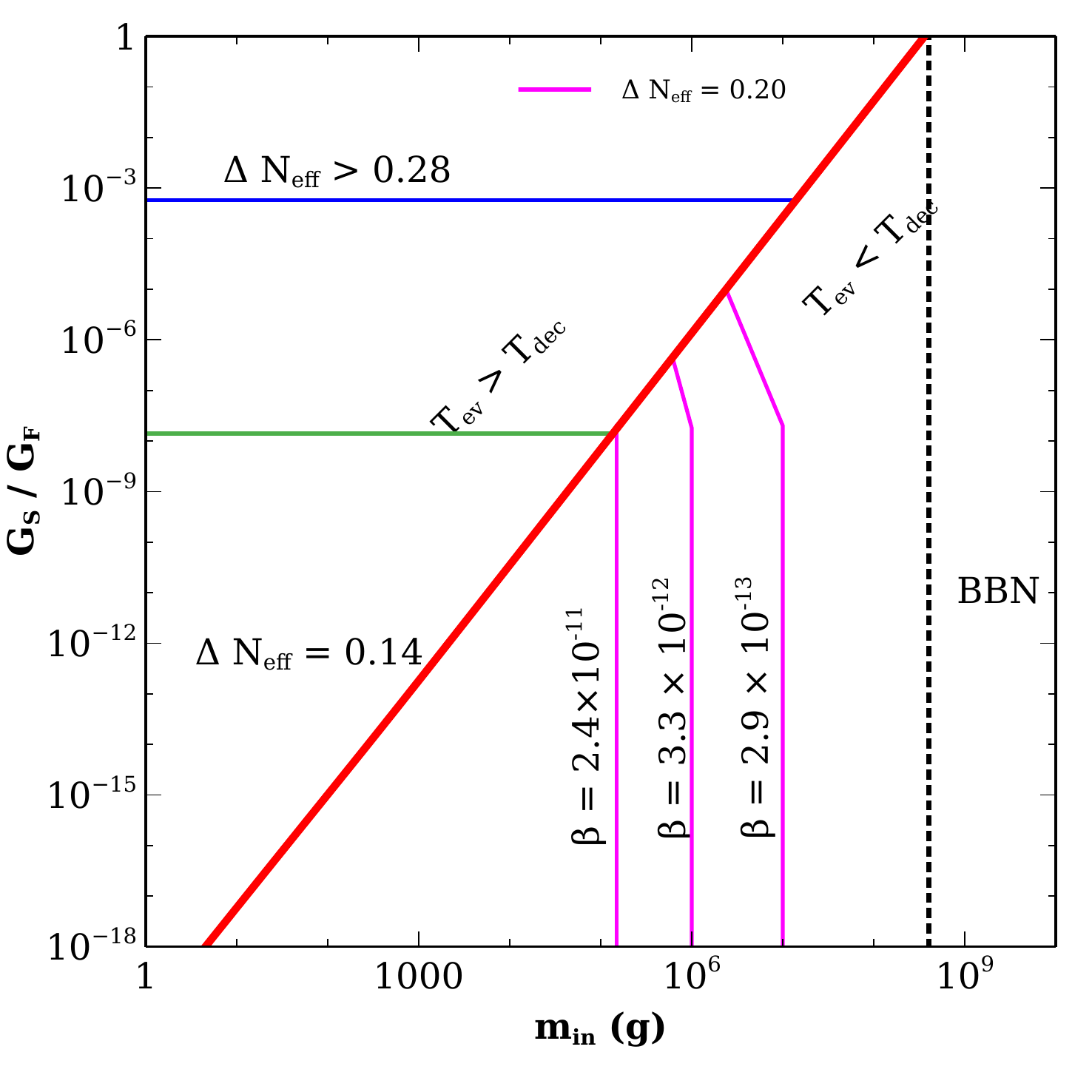}
\includegraphics[height=8cm,width=8.0cm,angle=0]{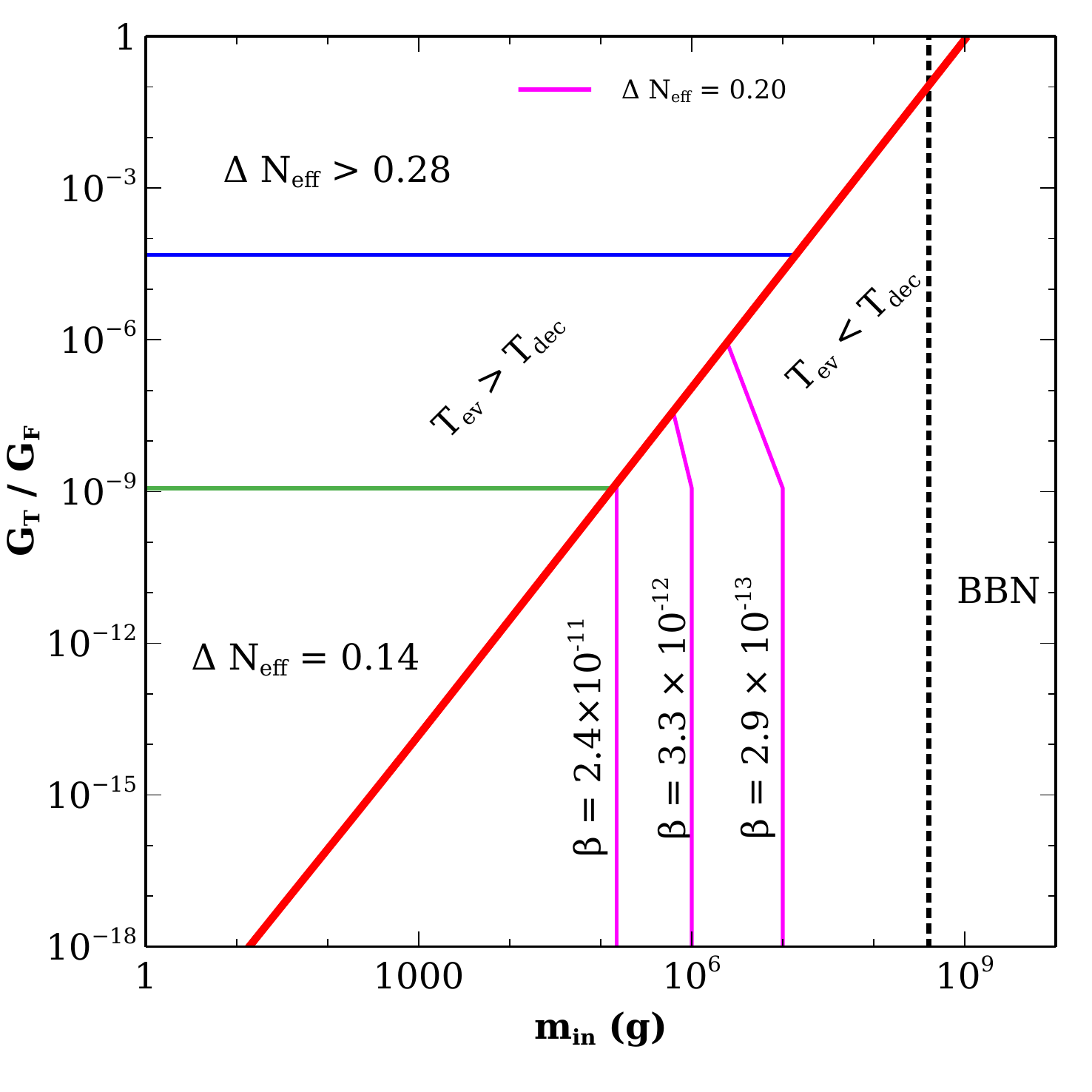}
\caption{${\rm \Delta N_{\rm eff}}$ contours in the plane of initial PBH mass versus the effective neutrino coupling for the case of radiation domination. In the left panel, $G_{T} = \Tilde{G}_{S} = G_{V} = 0$ with non-zero $G_{S}$, whereas in the right panel, we consider $G_{S} = \Tilde{G}_{S} = G_{V} = 0$ with non-zero $G_{T}$.The black vertical dashed line indicates the upper bound on PBH mass from BBN limit.}
\label{fig:Radiation_Dominated}
\end{figure}

With this, we redraw the Fig. \ref{fig:Mass_vs_tdec} by replacing $T_{\rm dec}$ in the y-axis with the equivalent $G_{\rm eff}$. The resulting plots are shown in Fig. \ref{fig:PBH_dominated} and Fig. \ref{fig:Radiation_Dominated} for PBH and radiation domination respectively. In the left panel plot of Fig. \ref{fig:PBH_dominated}, the coupling $G_{S}$ is non-zero while the rest (i.e. $G_{T}, \Tilde{G_{S}}$ and $G_{V}$) are zero. In the right panel plot, $G_{T}$ is non-zero and the rest are set to zero. For $G_{\rm eff} \gtrsim 10^{-3} G_{\rm F}$ with $G_F$ being the Fermi coupling, the decoupling temperature of $\nu_R$ turns to be $\lesssim 500$ MeV and ${\rm \Delta N_{\rm eff}}$ is more than 0.28. Similarly, below a certain value of coupling, $G_{\rm eff} \sim 2\times 10^{-7} G_{\rm F}$, the contribution to ${\rm \Delta N_{\rm eff}}$ saturates at a value of $0.14$. The corresponding parameter space is shown for the case of radiation domination in Fig. \ref{fig:Radiation_Dominated}, which can be understood in a way analogous to the right panel plot of Fig. \ref{fig:Mass_vs_tdec} discussed earlier.

Now, in order to check  whether the right-handed neutrinos produced after PBH evaporation enters into thermal equilibrium with the SM bath, we calculate the interaction of $\nu_{L}$ present in the bath and non-thermal $\nu_{R}$ produced from PBH evaporation. Following \cite{Cheek:2021cfe}, the thermal averaged cross-section for two massless species having different temperatures is found to be
%\begin{eqnarray}
  %  \frac{n^{\rm eq}_{\nu_{L}}(T_{\rm ev}) \langle \sigma v \rangle_{T_{\rm ev}T_{\rm BH}}}{H(T_{\rm ev})}.
%\end{eqnarray}
%Here $n^{\rm eq}_{\nu_{L}}$ is the equilibrium number density of the target particle i.e. $\nu_{L}$. The PBH generated $\nu_R$ has a black hole temperature $T_{\rm BH}$ given by Eq. \eqref{eq:T_BH} ( For the PBH mass range of our interest, $T_{\rm BH} \gg T_{\rm ev}$ ). 
\begin{eqnarray}
\langle \sigma v \rangle_{T_{\rm ev}T_{\rm BH}} = \frac{1}{32 (T_{\rm ev}T_{\rm BH})^{5/2} } \int_{0}^{\infty} \sigma s^{3/2} K_{1}\left(\frac{\sqrt{s}}{\sqrt{T_{\rm ev}T_{\rm BH}}}\right) ds,
\end{eqnarray}
with $K_1$ being the modified Bessel's function of order 1. For $\sigma = \frac{1}{4 \pi} \, s \, G^{2}_{\rm eff}$, this gives
\begin{eqnarray}
    \left<\sigma v \right>_{T_{\rm ev}T_{\rm BH}} = \frac{6}{\pi} \, G_{\rm eff}^2 \, T_{\rm ev} T_{\rm BH}
\end{eqnarray}
Comparing the interaction rate corresponding to the above cross-section with the Hubble  leads to the thermalisation condition 
\begin{eqnarray} \label{con1}
    \frac{n^{\rm eq}_{\nu_{L}}(T_{\rm ev}) \langle \sigma v \rangle_{T_{\rm ev}T_{\rm BH}}}{H(T_{\rm ev})} = C \frac{6}{\pi} \, G_{\rm eff}^2 \, T_{\rm ev}^2 T_{\rm BH} =1\,,
\end{eqnarray}
where $C = \frac{{}\frac{3}{4} \zeta(3)\times 2 \times \sqrt{45} \times M_{P}}{\pi^2 \sqrt{4 \pi^3 g_{*}(T_{\rm ev})}}$.

\begin{figure}[h!]
\includegraphics[height=7cm,width=8.0cm,angle=0]{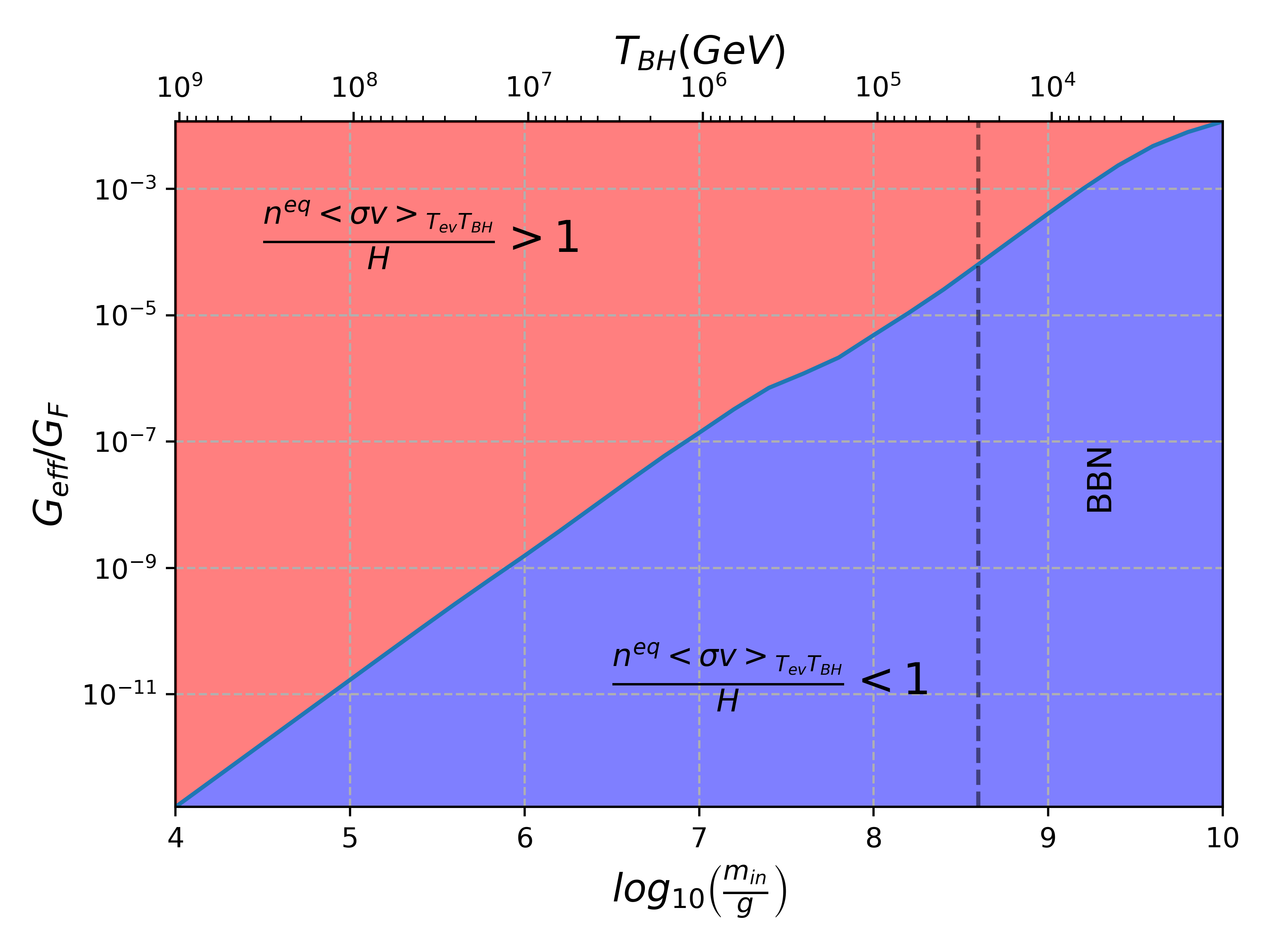}
\includegraphics[height=7cm,width=8.0cm,angle=0]{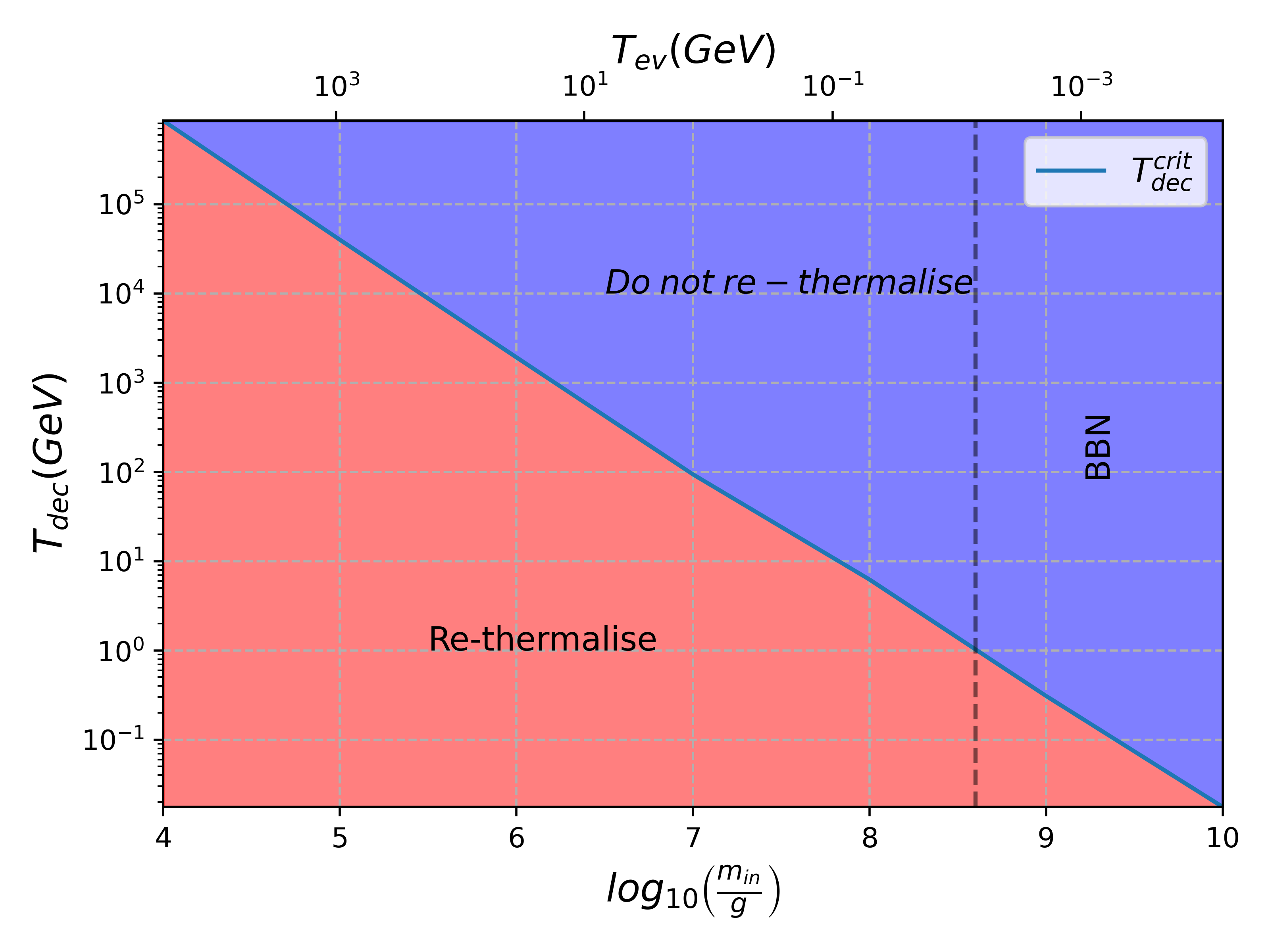}
\caption{\textit{Left panel}: $m_{\rm in}$ versus $G_{\rm eff}/G_{\rm F}$ parameter space showing the possibility of re-thermalisation. The red (blue) coloured region denotes the parameter space for which the PBH generated $\nu_{R}$ can (can not) come into thermal equilibrium. \textit{Right panel}: $m_{\rm in}$ versus $T_{\rm dec}$ parameter space showing the possibility of re-thermalisation. The red and blue coloured regions have the same meaning as the ones in left panel plot. The vertical dashed line indicates the upper bound on PBH mass from BBN limit.}
\label{fig:thermalisation_PBH_neutrino}
\end{figure} 
Both $T_{\rm ev}$ and $T_{\rm BH}$ are functions of $m_{\rm in}$ as can be seen from Eq.\eqref{eq:Tev} and Eq. \eqref{eq:T_BH}. In the left panel of  Fig. \ref{fig:thermalisation_PBH_neutrino}, we show the parameter space in the $m_{\rm in} - G_{\rm eff}$ plane, where re-thermalisation of the PBH generated $\nu_R$ takes place (red-shaded), along with the region which does not lead to such re-thermalisation (blue-shaded).  We can see that for a fixed PBH mass, if $G_{\rm eff}$ is smaller than a particular critical value, the non-thermally generated $\nu_{R}$ from PBH never thermalise again with the bath. A smaller value of $G_{\rm eff}$ also implies a higher initial decoupling temperature of thermal $\nu_R$. Hence, it is also possible to relate this critical value of $G_{\rm eff}$ below which re-thermalisation does not occur,  with the initial decoupling temperature of $\nu_{R}$. In the right panel of Fig. \ref{fig:thermalisation_PBH_neutrino}, we show the parameter space in the $m_{\rm in} - T_{\rm dec}$ plane, separating the region of re-thermalisation from the rest. For a particular PBH mass, a decoupling temperature higher than $T^{\rm crit}_{\rm dec}$ leads to the situation where $\nu_{R}$ from PBH will never re-thermalise with the bath \footnote{Considering a radiation dominated universe during decoupling, the value of this critical decoupling temperature is found to be 
\begin{align}
    T^{\rm crit}_{\rm dec} \simeq \left({T_{\rm ev}^2 T_{\rm BH}}\right){^{1/3}}.
\end{align}.}. 

 %The decoupling temperature of $\nu_{R}$ from the bath before the evaporation of PBH, can be estimated from the following condition 
%\begin{eqnarray} \label{con2}
 %   1 = \frac{n^{\rm eq}_{\nu_{L}}(T_{\rm dec}) \langle \sigma v \rangle_{T_{\rm dec}}}{H(T_{\rm dec})} = \frac{n^{\rm eq}_{\nu_{L}}(T_{\rm dec}) \frac{\pi}{6} \, G_{\rm eff}^2 \, T^{2}_{\rm dec}} {H(T_{\rm dec})} = C \frac{6}{\pi} \, G_{\rm eff}^2 \, T_{\rm dec}^3\,,
%\end{eqnarray}
%where the last equality holds for radiation domination. If this is true, it is possible to connect Eq. \eqref{con1} and Eq. \eqref{con2}. This is done if right side of figure \ref{fig:thermalisation_PBH_neutrino}. Thus for a particular PBH mass, a decoupling temperature higher than $T^{critical}_{dec}$ leads to the situation where $\nu_{R}$ from PBH will never thermalised with the bath. (\textcolor{red}{Last few lines are not clear})

To summarise, Fig. \ref{fig:thermalisation_PBH_neutrino} shows the constraints on the parameter space for which the $\nu_{R}$ produced from PBH evaporation will never re-thermalise. Once, we identify the parameter space where re-thermalisation occurs, the next step is to find the new ${\rm \Delta N_{\rm eff}}$ for these regions. However, this requires a careful treatment of complete Boltzmann equations to track the phase-space distribution function of evaporated $\nu_{R}$, which is numerically expensive and out of the scope of the current work. Here, we assume that the new decoupling temperature of $\nu_{R}$ after re-thermalisation is  $ \sim T_{\rm ev}$ for a particular PBH mass. This can be justified by the fact that PBH generated $\nu_R$ having a large initial energy and a sufficiently large interaction rate with the bath responsible for re-thermalisation, also suffer from instant dissipation of the energy to the bath. Due to this instant dissipation of energy, $\nu_R$ can not maintain equilibrium for a longer period after re-thermalisation leading to decoupling of $\nu_R$ once again at a temperature very close to $T_{\rm ev}$.

\begin{figure}[h!]
\includegraphics[height=6cm,width=8.0cm,angle=0]{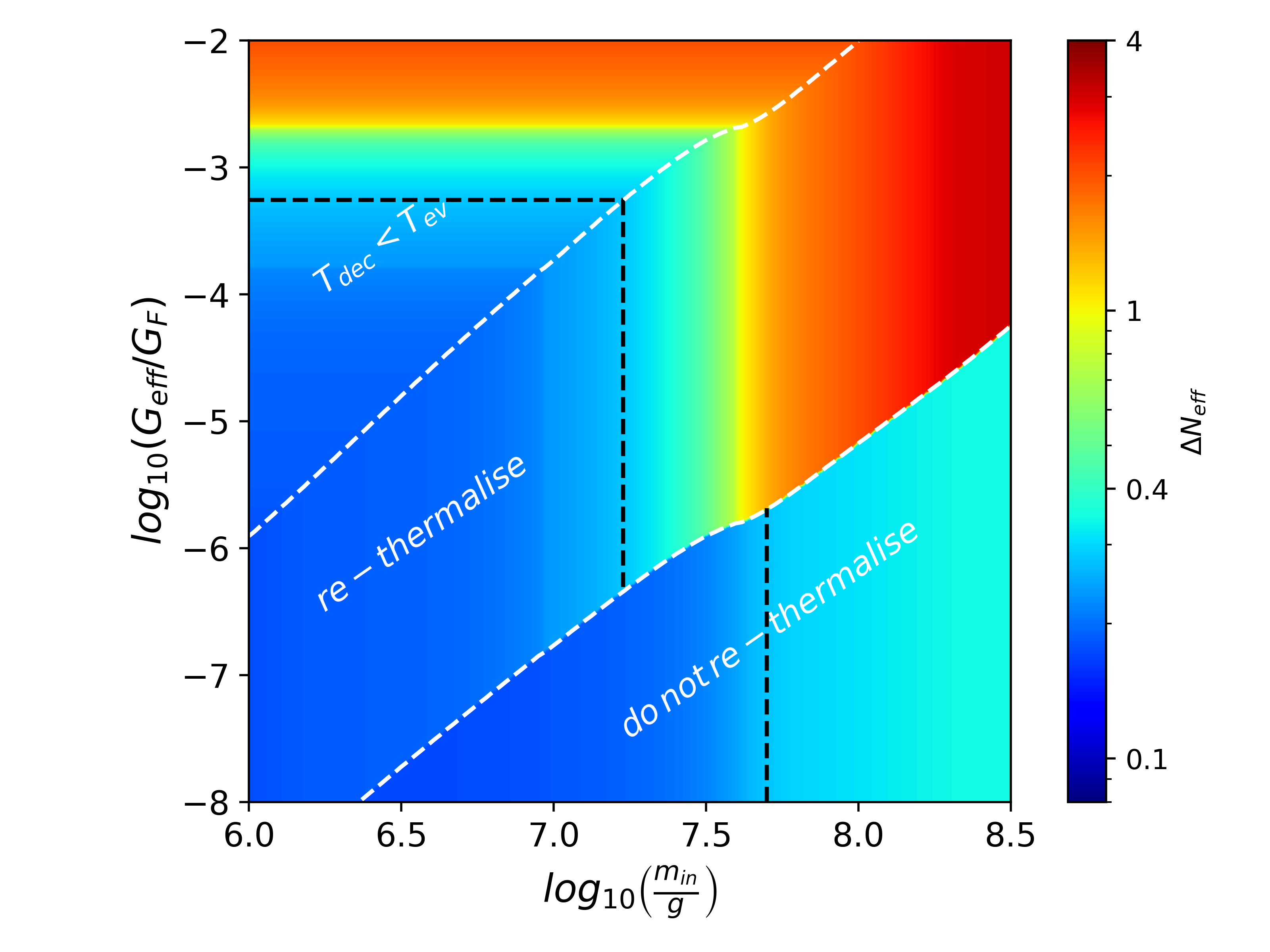}
\includegraphics[height=6cm,width=8.0cm,angle=0]{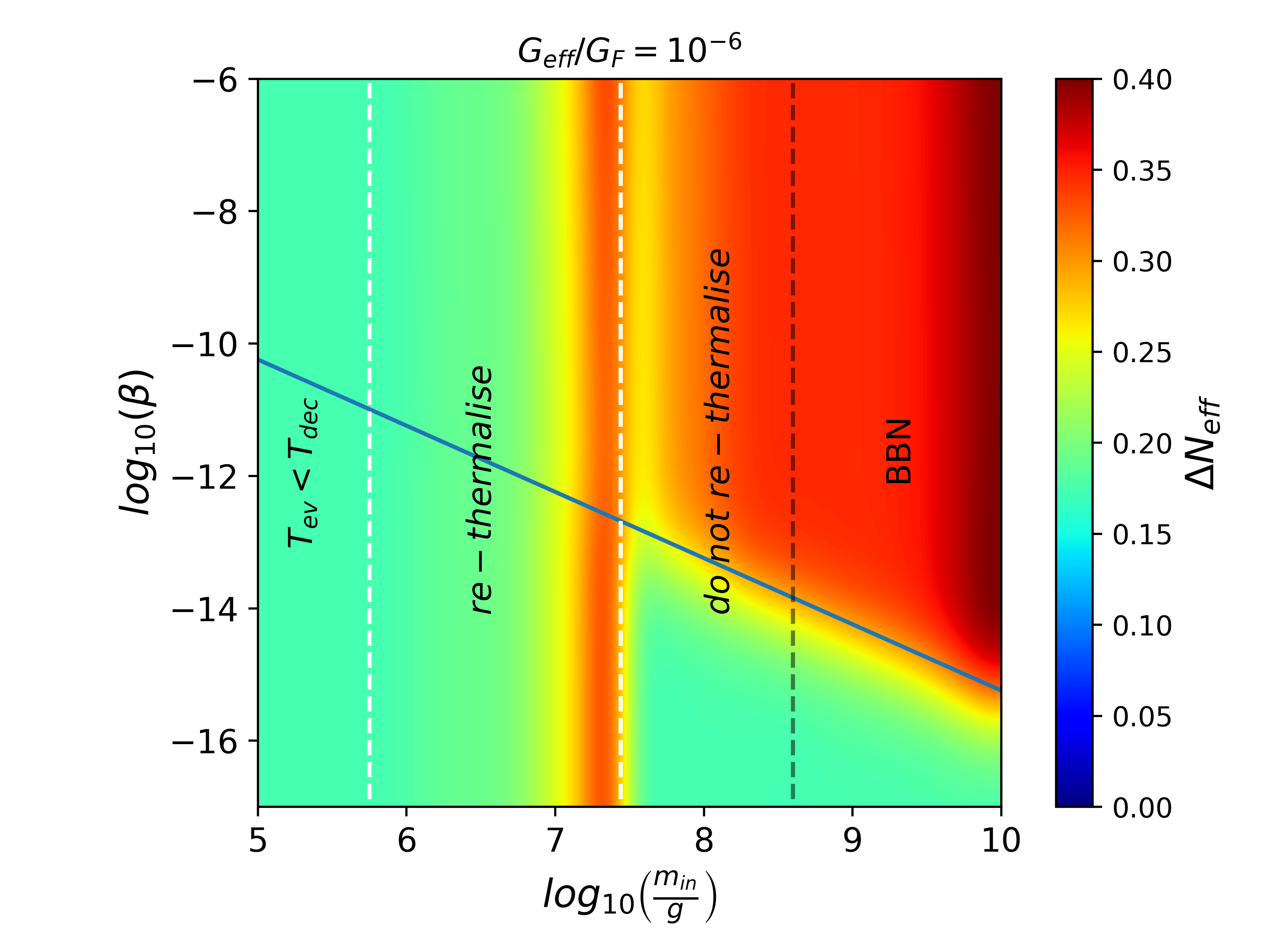}
\caption{\textit{Left panel}: Variation of ${\rm \Delta N_{\rm eff}}$ in $m_{\rm in}$-$G_{\rm eff}/G_{\rm F}$ plane for three different regions where (i) $ T_{\rm dec} < T_{\rm ev}$, (ii) PBH generated $\nu_R$ re-thermalise and (iii) PBH generated $\nu_R$ do not re-thermalise, separated by white dashed line. The dashed black coloured line distinguishes the regions with ${\rm \Delta N_{\rm eff}}>0.28$ and ${\rm \Delta N_{\rm eff}}<0.28$. \textit{Right panel:} Variation of ${\rm \Delta N_{\rm eff}}$ in the $m_{\rm in}$-$\beta$ plane for a particular $G_{\rm eff}/G_{\rm F}=10^{-6}$. A new constraint is obtained in the \textit{re-thermalised} region. The black vertical dashed line indicates the upper bound on PBH mass from BBN limit.}   
\label{fig:rethermalisation_Neff}
\end{figure}

%\sjd{Can we write a line here justifying this assumption?}
%This can be justified from the fact that after re-thermalisation, $\nu_{R}$ try to distribute its energy density. As a results, its temperature decreases and eventually reaches $T_{\rm ev}$. 
With this, it is possible to find new constraint on $G_{\rm eff}$ and $m_{\rm in}$. In the left panel of Fig. \ref{fig:rethermalisation_Neff}, constraints are shown in $m_{\rm in}$-$G_{\rm eff}$ plane. The black dashed lines denote ${\rm \Delta N_{\rm eff}} = 0.28$. Taking into account re-thermalisation, it is possible to obtain a stronger constraint on $G_{\rm eff}/G_{\rm F}$. In the region, $T_{\rm dec}<T_{\rm ev}$, ${\rm \Delta N_{\rm eff}} > 0.28$ implies $G_{\rm eff}/G_{\rm F} \gtrsim 5 \times 10^{-4}$. However, in the re-thermalisation region, $G_{\rm eff}/G_{\rm F}$ up to $\sim 5 \times 10^{-7}$ can be excluded, for $m_{\rm in}\gtrsim 10^{7.2}$ g. In the bottom triangular region, which shows the parameter space where PBH generated $\nu_{R}$ do not re-thermalise, the results correspond to the one discussed in Fig. \ref{fig:exact}. In this region, we have shown the ${\rm \Delta N_{\rm eff}}$ values  for a particular value of $\beta = 10^{-8}$. In the right panel of Fig. \ref{fig:rethermalisation_Neff}, we show the constraint in the $m_{\rm in}$ versus $\beta$ plane for a particular $G_{\rm eff}/G_{\rm F}=10^{-6}$. Here, the region between the white dashed lines shows the parameter space leading to re-thermalisation. As expected from the left plot, here a stronger constraint on PBH mass can be seen.

%However, as mentioned earlier, assumptions have been made in obtaining the results. We assume that all the particles from PBH have same temperature, i.e. $T_{BH}$ (or more precisely $T^{ini}_{BH}$). But in reality, as the mass of PBH decreases, the corresponding temperature should increase, which we have neglected. The next assumption is that while finding the thermal average cross section, the $\nu_{R}$ from PBH have maxwell-boltzmann distribution. Although this is not true, the actually distribution function can still be approximated using maxwell-boltzmann distribution as seen from figure . These two assumptions are for both the plots in \ref{fig:thermalisation_PBH_neutrino}. Finally, there is one more assumption in the right hand side plot, i.e. the universe should be radiation dominated at the time of $\nu_{R}$ decoupling. If the universe is PBH dominated at $\nu_{R}$ decoupling, we will get a higher critical decoupling temperature $T^{critical}_{dec}$ for a particular PBH mass. 

\section{Gravitational waves from PBH density perturbations}
\label{sec:sec5}
PBH can be involved in the generation of gravitational waves in several ways \cite{Anantua:2008am, Zagorac:2019ekv, Saito:2008jc}. In this work, we focus on the gravitational waves induced by the density fluctuations of PBH after their formation \cite{Papanikolaou:2020qtd, Domenech:2020ssp, Domenech:2021wkk}. This is primarily because for the PBH mass range we are working with, the GW spectrum produced through this route can be within the sensitivity of near-future GW experiments. Moreover, these induced gravitational waves are independent of the formation mechanism of PBH.

After the formation of PBH, they are distributed randomly in space following Poissonian statistics \cite{Papanikolaou:2020qtd}. These inhomogeneities in the distribution of PBH induce curvature perturbations when PBH begin to dominate the energy density of the universe, which at second order can source gravitational waves. The amplitude of these gravitational waves is further enhanced during the evaporation of PBH. The dominant contribution to the present-day GW amplitude can be written as\footnote{The amplitude of the induced GW spectrum is sensitive to the PBH mass distribution \cite{Papanikolaou:2022chm}. Here, we consider a monochromatic mass spectrum for simplicity, as mentioned earlier.} \cite{Domenech:2020ssp, Borah:2022iym, Barman:2022pdo}
\begin{equation}
    \ogw(t_0,f)\simeq \ogw^{\rm peak}\left(\frac{f}{f^{\rm peak}}\right)^{11/3}\Theta
\left(f^{\rm peak}-f\right),\label{eqn:omgw}
\end{equation}
where $ \ogw^{\rm peak}$ indicates the peak amplitude and is given by
\begin{equation}
    \ogw^{\rm peak}\simeq 2\times 10^{-6} \left(\frac{\beta}{10^{-8}}\right)^{16/3}\left(\frac{m_{\text{in}}}{10^7 \rm g}\right)^{34/9}.\label{eqn:omgpeak}
\end{equation}
Now, for length scales smaller than the mean separation between PBH, the assumption of PBH as a continuous fluid ceases to hold true. This imposes an ultraviolet cutoff to the GW spectrum, with $f^{\rm peak}$ corresponding to comoving scales representing the mean separation between PBH. The peak frequency is found to be
\begin{equation}
    f^{\rm peak}\simeq 1.7\times 10^3\,{\rm Hz}\,\left(\frac{m_{\text{in}}}{10^4 \rm g}\right)^{-5/6}.\label{eqn:fpk}
\end{equation}

\begin{figure}[h!]
\includegraphics[height=6cm,width=8.0cm,angle=0]{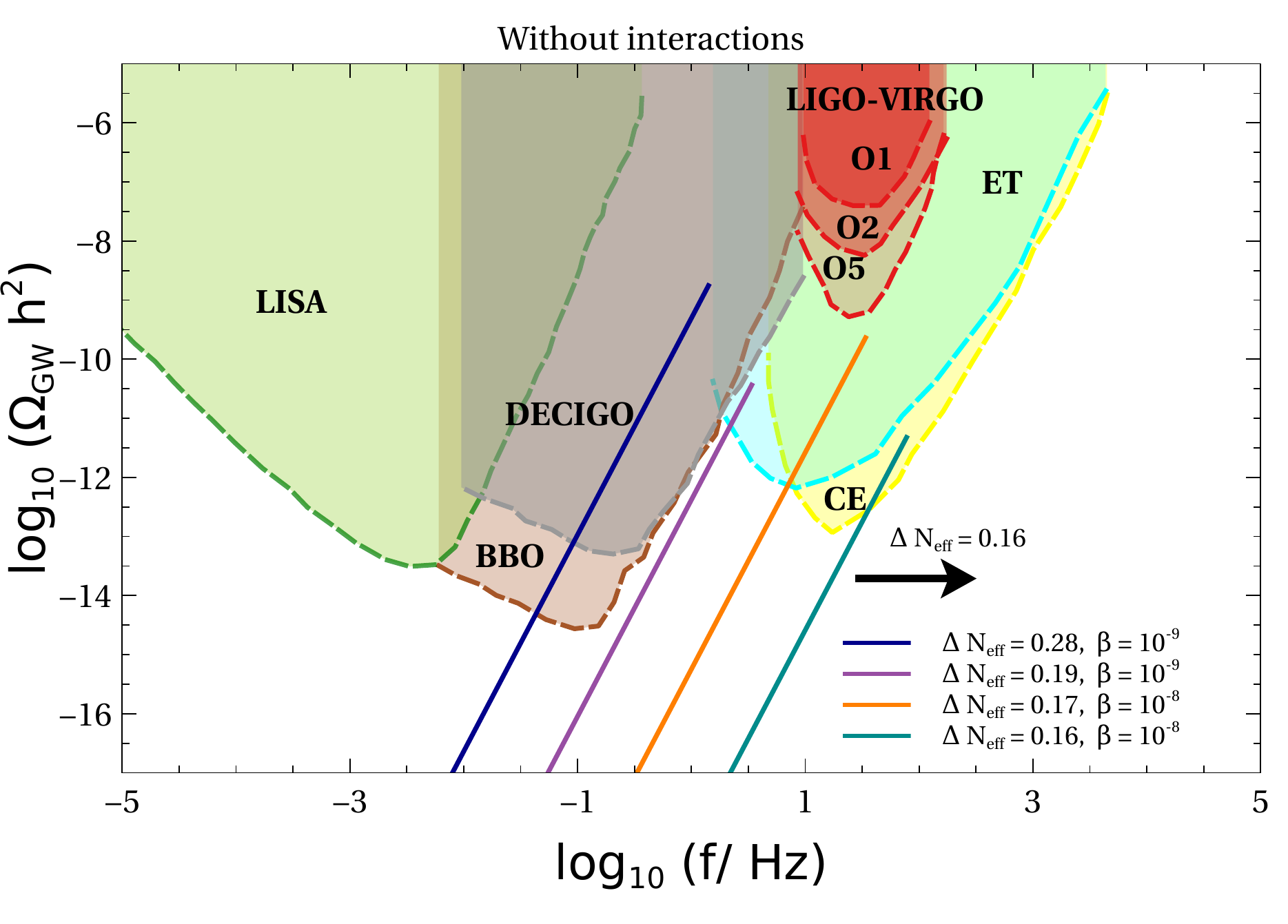}
\includegraphics[height=6cm,width=8.0cm,angle=0]{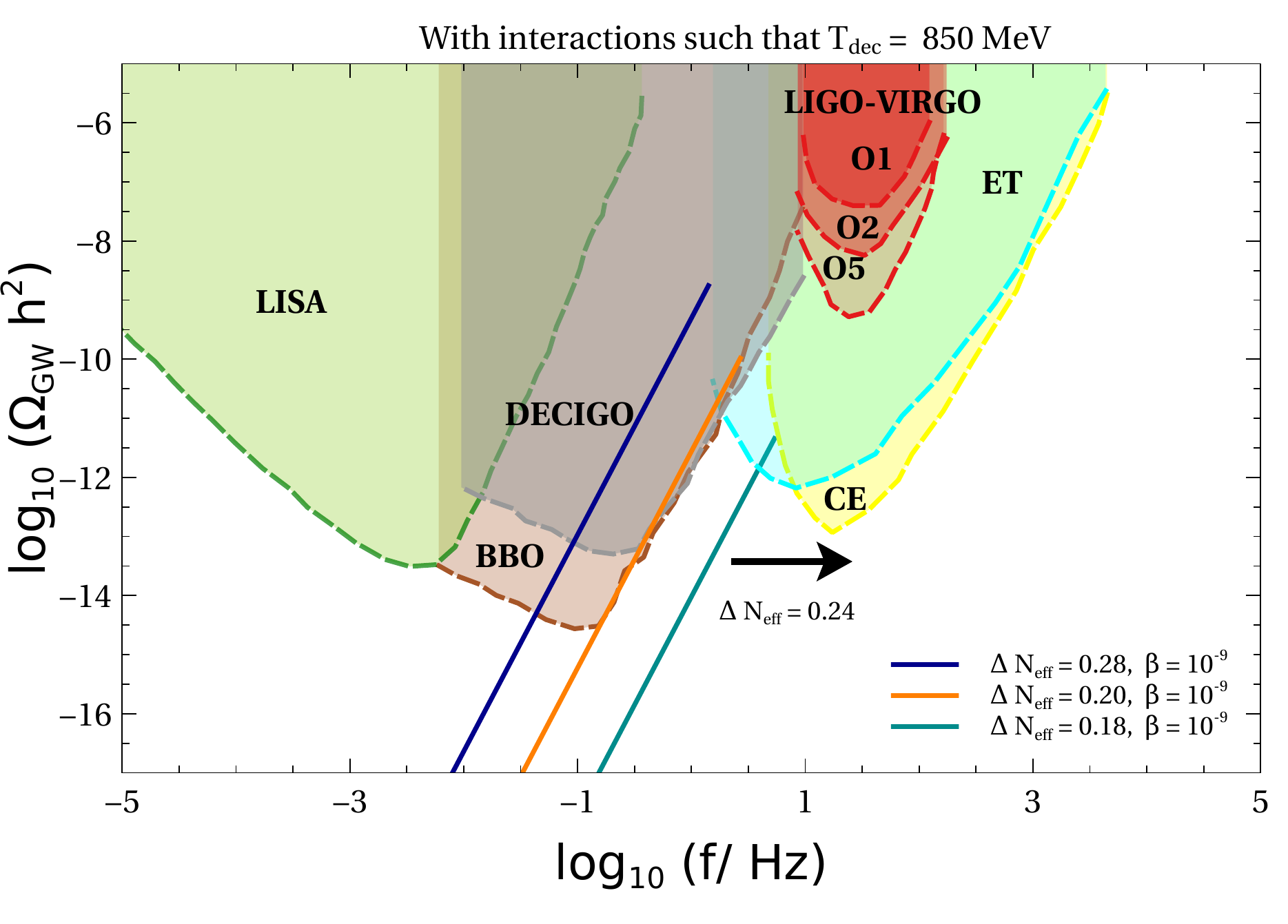}
\caption{GW spectra from PBH density fluctuations for different values of ${\rm \Delta N_{\rm eff}}$, without extra interactions (left panel) and in the presence of extra interactions (right panel) of right-handed neutrinos. The various shaded regions indicate the future sensitivities and current bounds of GW detectors which includes LISA, BBO, DECIGO, LIGO-VIRGO, ET, CE. The arrows indicate a constant ${\rm \Delta N_{\rm eff}}$ beyond a particular frequency (see text).}
\label{fig:igw_neff}
\end{figure}

Since GW behave like radiation, they can contribute to extra relativistic degrees of freedom during BBN. This gives an upper bound on $\beta$, depending on PBH mass, which is given by \cite{Domenech:2020ssp}
\begin{equation}
    \beta \lesssim 1.1 \times 10^{-6} \left(\frac{m_{\rm in}}{10^{4}g}\right)^{-17/24}\,.\label{eq:betmax}
\end{equation}
This upper bound on $\beta$ for ultralight PBH is stronger than other bounds, obtained for eg. in Ref. \cite{Papanikolaou:2020qtd} to avoid the backreaction problem.  

Now, if PBH play a role in sourcing the extra relativistic degrees of freedom ${\rm \Delta N_{\rm eff}}$, then the peak amplitude and the peak frequency of the GW spectrum discussed above would depend on the value of ${\rm \Delta  N_{\rm eff}}$, since $m_{\rm in}$ in Eq. \eqref{eqn:omgpeak}, \eqref{eqn:fpk} is connected to ${\rm \Delta  N_{\rm eff}}$, as we have already seen in earlier discussions. Hence, observation of such a GW spectrum would provide a complementary probe to ${\rm \Delta  N_{\rm eff}}$ observations at CMB experiments. In the left panel of Fig. \ref{fig:igw_neff}, we show the GW spectrum arising from PBH density fluctuations, for different values of ${\rm \Delta  N_{\rm eff}}$, in the absence of extra interactions of the right-handed neutrinos. Note that such a GW spectrum is absent for the radiation-dominated case. For PBH domination, a higher value of ${\rm \Delta  N_{\rm eff}}$ corresponds to a higher value of $m_{\rm in}$ (cf. Fig. \ref{fig:neff_comparison}), which shifts the peak frequency given by Eq. \eqref{eqn:fpk} to lower values. Any spectral line to the right of the one coloured in dark cyan, would correspond to the same value of ${\rm \Delta  N_{\rm eff}} \simeq 0.16$, since PBH masses in that range evaporate above the electroweak scale, giving the same value of ${\rm \Delta  N_{\rm eff}}$ (see Fig. \ref{fig:neff_comparison}). Here, $\beta$ remains a free parameter as for PBH domination, ${\rm \Delta  N_{\rm eff}}$ is independent of $\beta$. In the right panel of Fig. \ref{fig:igw_neff}, we show the GW spectra in the presence of extra interactions of right-handed neutrinos, with the strength of the interactions taken such that the thermal $\nu_R$ decouple at a temperature of $\sim 850$ MeV. This case corresponds to the bottom right panel of Fig. \ref{fig:exact}. Here, any spectral line to the right of the one coloured in dark cyan, would correspond to the same value of ${\rm \Delta  N_{\rm eff}} = 0.24$, since PBH in this mass range evaporate at a temperature higher than $850$ MeV. Thus, the thermal contribution would dominate as discussed earlier, contributing to a total ${\rm \Delta N_{\rm eff}}$ of $0.24$, similar to the bottom right panel of Fig. \ref{fig:exact}. In both the plots shown in Fig. \ref{fig:igw_neff}, the experimental sensitivities of relevant GW detectors namely, LISA\,\cite{2017arXiv170200786A}, DECIGO \cite{Kawamura:2006up}, BBO\,\cite{Yagi:2011wg}, ET\,\cite{Punturo_2010}, CE\,\cite{LIGOScientific:2016wof} and LIGO-VIRGO \cite{LIGOScientific:2014pky} are shown.

\begin{figure}
    \centering
    \includegraphics[scale=0.6]{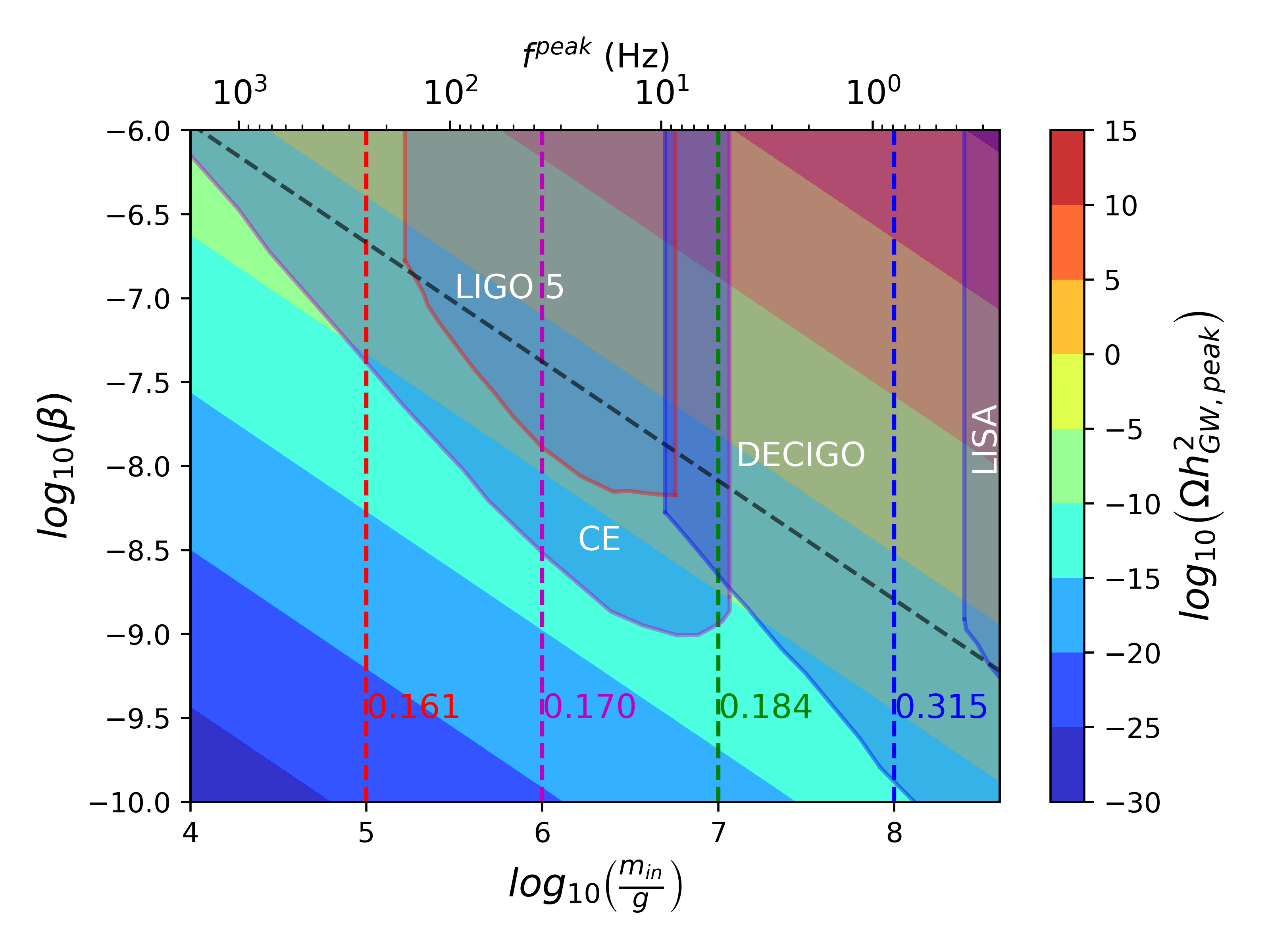}
    \caption{Contours of peak amplitude of GW induced by PBH density perturbations. The corresponding ${\rm \Delta N_{\rm eff}}$ due to PBH generated $\nu_R$ are shown assuming no (or very feeble) $\nu_R$-SM interactions. The black dashed line indicates the upper bound on $\beta$ (cf. Eq. \eqref{eq:betmax}).}
    \label{fig:gw_beta_mass}
\end{figure}

As we have seen already, the value of ${\rm \Delta N_{\rm eff}}$ is independent of $\beta$ for PBH domination. Thus, future observations of ${\rm N_{eff}}$ from CMB observations will not be able to tell us anything about $\beta$ provided $\beta > \beta_{\rm crit}$. However, the value of $\beta$ can be inferred from future GW detection with the spectral shape similar to the one generated by PBH density perturbations, since the peak of the GW amplitude is determined by $\beta$. In Fig. \ref{fig:gw_beta_mass}, we show the contours of the peak amplitude of GW in the $m_{\rm in}-\beta$ plane, indicating the values of $f_{\text{peak}}$ and ${\rm \Delta N_{\rm eff}}$, considering no extra interactions of $\nu_R$ with the SM. The future sensitivities of several GW experiments are also shown in the same figure. In the presence of extra interactions, the values of ${\rm \Delta N_{\text{eff}}}$ would be different as discussed above (cf. Fig. \ref{fig:igw_neff}). ${\rm \Delta N_{\rm eff}}$ values would also be different for other light species. Finally, we summarise the GW and CMB complementarities in table \ref{BP1} and \ref{BP2}
by choosing a few benchmark points. PBH mass and initial energy fraction are fixed at benchmark values and the corresponding implications for ${\rm \Delta N_{eff}}$ and GW observations for different types of dark radiation namely, $\nu_R$, Goldstone boson, massless gauge boson are shown in table \ref{BP1} and \ref{BP2}. Note that we work with $\beta$ values consistent with the upper bound given by Eq. \eqref{eq:betmax}, and higher than the critical value $\beta_{\text{crit}}$(cf. Eq. \eqref{eq:betacr}) required for PBH to dominate. While in table \ref{BP1}, no DR-SM interactions are assumed, in table \ref{BP2}, thermalised DR is considered with specific decoupling temperatures. While BP1 and BP2 in both the cases remain within future sensitivities, BP3 for some specific type of DR is already ruled out by Planck $2\sigma$ limits.
\begin{table}[]
    \centering
    \caption{Without DR-SM interaction}
    \begin{tabular}{|c|c|c|c|c|c|c|c|c|c|}
    \hline
    \multirow{2}{*}{BP} & $m_{\rm in}$ & \multirow{2}{*}{$\beta$} & \multicolumn{3}{|c|}{${\rm \Delta N_{\rm eff}}$} & \multicolumn{3}{|c|}{CMB experiment ($2\sigma$)} & \multirow{2}{*}{GW experiment}\\  \cline{4-9} \multirow{2}{*}{} & (g) & \multirow{2}{*}{} &  $ \nu_{R}$ & GB & MGB & $\nu_{R}$ & GB & MGB & \multirow{2}{*}{} \\ \hline 
    BP1 & $10^{6}$ & $1.5 \times 10^{-8}$ & $0.167$ & $0.052$ & $0.023$ & CMB-S4 & CMB-HD & None & CE, ET, LIGO-VIRGO  \\
    BP2 & $10^{7}$ & $6 \times 10^{-9}$ & $0.183$ & $0.056$ & $0.024$ & CMB-S4 & CMB-HD & None & CE, ET, DECIGO, BBO \\
    BP3 & $10^{8}$ & $6 \times 10^{-10}$ & $0.313$ & $0.096$ & $0.042$ & Planck & CMB-S4 & CMB-HD & DECIGO, BBO \\
    \hline
    \end{tabular}
    \label{BP1}
\end{table}

\begin{table}[]
    \centering
    \caption{With DR-SM interaction}
    \begin{tabular}{|c|c|c|c|c|c|c|c|c|c|c|}
    \hline
    \multirow{2}{*}{BP} & {$m_{\rm in}$} & \multirow{2}{*}{$\beta$} & {$T_{\rm dec}$} & \multicolumn{3}{|c|}{${\rm \Delta N_{\rm eff}}$} & \multicolumn{3}{|c|}{CMB experiment ($2\sigma$)} & \multirow{2}{*}{GW experiment} \\\cline{5-10} \multirow{2}{*}{} & {(g)} & \multirow{2}{*}{} & {(GeV)} & $\nu_{R}$ & GB & MGB & $\nu_{R}$ & GB & MGB & \multirow{2}{*}{} \\ \hline 
    BP1 & $10^{6}$ & $1.5 \times 10^{-8}$ & $10$ & $0.188$ & $0.035$ & $0.072$ & CMB-S4 & CMB-HD & CMB-S4 & CE, ET, LIGO-VIRGO \\
    BP2 & $10^{7}$ & $6 \times 10^{-9}$ & $0.8$ & $0.242$ & $0.046$ & $0.092$ & CMB-S4 & CMB-HD & CMB-S4 & CE, ET, DECIGO, BBO \\
    BP3 & $10^{8}$ & $6 \times 10^{-10}$ & $0.01$ & $3.02$ & $0.575$ & $1.15$ & Planck & Planck & Planck & DECIGO, BBO \\
    \hline
    \end{tabular}
    \label{BP2}
\end{table}

\section{Conclusion}
\label{sec:sec6}
We revisit the prospects of generating dark radiation from primordial black holes in the early universe, by considering sizeable DR-SM interactions. We first reproduce the results by considering PBH and thermal interactions separately and the consequences for ${\rm \Delta N_{eff}}$ observations at CMB experiments. We then consider the hybrid scenario with both thermal interactions and PBH to be source of DR and put new constraints on the PBH parameters as well as DR-SM interactions from the requirement of satisfying BBN and CMB limits on ${\rm \Delta N_{eff}}$. Compared to the scenario with no DR-SM interactions discussed in earlier works, our present scenario puts tighter constraints on the PBH parameter space namely, initial PBH mass and energy density. 

We also find an interesting region of parameter space involving PBH as well as DR-SM interactions where DR initially decouples from the bath followed by PBH evaporation leading to re-thermalisation of DR with the SM bath. This leads to a lower decoupling temperature of DR enhancing the ${\rm \Delta N_{eff}}$ and putting new constraints in the PBH parameter space. Though we have considered a specific type of DR namely, light Dirac neutrinos for the numerical analysis, the generic conclusions are applicable to any thermalised DR.

We also find the gravitational wave complementarity of this scenario by considering PBH density perturbations to be the source of such stochastic GW. In the ultra-light PBH window considered in our work, GW sourced from PBH this way not only remains within current and planned experiment's sensitivity but also remains independent of PBH formation mechanism. We show the complementarity between GW and CMB observations for different types of DR with and without DR-SM interactions. Interestingly, some of the benchmark points can lead to GW peak amplitude, frequency in LIGO-VIRGO ballpark while keeping ${\rm \Delta N_{eff}}$ within the reach future CMB experiments like CMB-S4. On the contrary, some part of the parameter space can keep GW prospects within future experiment's reach while saturating Planck 2018 limits on ${\rm \Delta N_{eff}}$. Such complementary detection prospects at CMB and GW experiments of thermalised dark radiation is particularly interesting due to limited prospects of detecting such light degrees of freedom at particle physics experiments.

Before we end, we briefly comment on the impact of Kerr PBH and a non-monochromatic mass function of PBH. First of all, considering Kerr PBH does not significantly alter DR contribution to $\Delta N_{\rm eff}$. This is true for DR with spin $0$, $1/2$ and $1$, but significant enhancement is observed for spin $2$ DR particles \cite{Hooper:2020evu, Masina:2021zpu, Cheek:2022dbx}. At the same time, our results on the analysis of GW spectrum would almost be unchanged. This is mainly because the evaporation temperature of a spinning BH changes only slightly compared to the non-spinning case \cite{Dong:2015yjs, Arbey:2019jmj}. On the other hand, $\Delta N_{\rm eff}$ can depend significantly on the mass distribution of PBH. In Ref. \cite{Cheek:2022mmy}, it was shown that the contribution to $\Delta N_{\rm eff}$ might substantially change for some choices of the mass spectrum, compared to the monochromatic case. Similarly, the amplitude of induced GW is also sensitive to the PBH mass spectrum \cite{Domenech:2021ztg, Papanikolaou:2022chm}. It would be interesting to study the combined effect in the situation where DR is produced both from thermal bath (due to sizeable DR-SM couplings) and from PBH with non-monochromatic mass distribution and non-zero spin distribution. We leave such a complete study for future works.

\section*{Acknowledgements}
%%%%%%%%%%%%%%%%
 ND would like to acknowledge Ministry of Education, Government of India for providing financial support for his research via the Prime Minister's Research Fellowship (PMRF) December 2021 scheme. SJD thanks Dibyendu Nanda for some useful discussions related to this project. The work of DB is supported by the Science and Engineering Research Board (SERB), Government of India grant MTR/2022/000575.
%\newpage
\appendix

\section{Non-standard interactions of dark radiation}\label{appendix:dirac_neutrino}
We consider the following effective operators for interactions among SM neutrinos and $\nu_R$ \cite{Luo:2020sho}.
\begin{eqnarray}
    -\mathcal{L} \supset G_{s} \Bar{\nu}_{L}\nu_{R}\Bar{\nu}_{L}\nu_{R} + G_{s}^{*}\Bar{\nu}_{R}\nu_{L}\Bar{\nu}_{R}\nu_{L} + G_{V} \Bar{\nu}_{L}\gamma^{\mu}\nu_{L} \Bar{\nu}_{R}\gamma_{\mu}\nu_{R} + \Tilde{G}_{S}\Bar{\nu}_{L}\nu_{R}\Bar{\nu}_{R}\nu_{L} \nonumber \\
    + G_{T} \Bar{\nu}_{L}\sigma^{\mu\nu}\nu_{R} \Bar{\nu}_{L}\sigma_{\mu\nu}\nu_{R} + G^{*}_{T} \Bar{\nu}_{R}\sigma^{\mu\nu}\nu_{L} \Bar{\nu}_{R}\sigma_{\mu\nu}\nu_{L},
\end{eqnarray}
where $G_{S}, \Tilde{G}_{S}, G_{V}$ and $G_{T}$ are effective coupling constants of respective 4-fermion operators. The relevant processes which can lead to thermalisation of $\nu_R$ are given by
\begin{eqnarray}
    \nu_{R} + \nu_{R} \leftrightarrow \nu_{R} + \nu_{R}, \nonumber \\
    \nu_{R} + \Bar{\nu}_{R} \leftrightarrow \nu_{L} + \Bar{\nu}_{L}, \nonumber \\
    \nu_{R} + \nu_{L} \leftrightarrow \nu_{R} + \nu_{L}, \nonumber \\
    \nu_{R} + \Bar{\nu}_{L} \leftrightarrow \nu_{R} + \Bar{\nu}_{L}, \nonumber \\
    \nu_{R} + \Bar{\nu}_{L} \leftrightarrow \Bar{\nu}_{R} + \nu_{L}.    
\end{eqnarray}

While we have focused on the case of right-handed neutrinos, interactions of other species of dark radiation with Standard Model can also be realised through effective operators. For example, coupling of Goldstone bosons ($\phi$) with the SM gauge sector at dimension-5 can be written as \cite{Baumann:2016wac}
    \begin{equation}
        \mathcal{L} \supset -\frac{1}{4}\frac{\phi}{\Lambda} \left( c_{1} B_{\mu\nu} \tilde{B}^{\mu\nu} + c_{2} W_{\mu\nu} \tilde{W}^{\mu\nu} + c_{3} G_{\mu\nu} \tilde{G}^{\mu\nu}\right),
    \end{equation}
    where $\{B_{\mu\nu},W_{\mu\nu},G_{\mu\nu}\}$ denotes the field strength tensor associated with SM gauge groups $\{U(1)_{Y}, SU(2)_{L}, SU(3)_{c}\}$, with $\{\tilde{B}^{\mu\nu},\tilde{W}^{\mu\nu},\tilde{G}^{\mu\nu}\}$ being their duals. Interaction of Goldstone bosons with the SM can also be realised through Yukawa couplings given by \cite{Baumann:2016wac}
    \begin{equation}
        \mathcal{L} \supset - \frac{\partial_{\mu}\phi}{\Lambda_{\psi}} \bar{\psi}_{1}\gamma^{\mu} (g^{ij}_{V} + g^{ij}_{A}\gamma^{5})\psi_{j}\,,
    \end{equation}
where $\psi$ denotes the SM fermions. Similar analysis performed in Section \ref{sec:sec4} for Dirac neutrinos can be carried out to constrain the couplings of other species of dark radiation.

\bibliographystyle{JHEP}
\bibliography{ref, ref1, ref2, ref3}

\end{document}